\documentclass[10pt,a4paper]{article}
\usepackage[utf8]{inputenc}
\usepackage{amsmath}
\usepackage{amsfonts}
\usepackage{amssymb}
\usepackage{graphicx}
\usepackage[left=2cm,right=3cm,top=2cm,bottom=2cm]{geometry}

\newcommand{\bk}{{k}}
\newcommand{\bx}{{x}}

\numberwithin{equation}{section}

\title{An introduction to the Zakharov equation for modelling deep water waves}
\author{Raphael Stuhlmeier\\School of Engineering, Computing \& Mathematics\\University of Plymouth\\PL4 8AA Plymouth, UK\\\texttt{raphael.stuhlmeier@plymouth.ac.uk}}
\date{}
\begin{document}

\maketitle

\begin{abstract}
The Hamiltonian formulation of the water wave problem due to Zakharov, and the reduced Zakharov equation derived therefrom, have great utility in understanding and modelling water waves. Here we set out to review the cubic Zakharov equation and its uses in understanding deterministic waves in deep water. The background of this equation is developed and several applications are explored. Chief among these is an understanding of dispersion corrections and the energy exchange among modes. It is hoped that readers will be motivated to explore this powerful reformulation of the cubically nonlinear water wave problem for themselves.
\end{abstract}

\section{Introduction}
The mathematical formalisation of fluid dynamics began with the work of Johann and Daniel Bernoulli and Jean le Rond d'Alembert in the first half of the 18th century. The completion of this work -- at least in the sense of formulating the equations of motion as partial differential equations -- was expounded by Leonhard Euler in two papers of 1752 and 1755. Having established fluid dynamics firmly as a discipline of classical mechanics, Euler was optimistic that ``all the theory of the motion of fluids has just been reduced to the solution of analytic formulas." The efforts of the last 250 years of fluid mechanics have shown that solving these analytic formulas is not so simple.

The inviscid flow equations derived by Euler contain a rich vastness of phenomena, but one of the most attractive is the behaviour of water waves. Their study adds an additional layer of complexity to the problem, in the form of an unknown water surface, which moves in space and time and must be determined together with the movement of the fluid beneath it. One natural approach to this complex problem would be to track the motion of individual particles, as is usually done in Newtonian mechanics. In this way a natural notion of kinetic and potential energy for each particle could be used to formulate a Lagrangian (or Hamiltonian) form of the equations of motion. The evident disadvantage is that such a formulation requires tracking very many particles, and the corresponding Lagrangian fluid mechanics is somewhat cumbersome. The more conventional approach, which deals with a fluid velocity field and forgets about the individual particle motions (which can anyhow be obtained retrospectively by integration) -- called Eulerian fluid mechanics -- has the advantage of simplicity and economy, and has traditionally displaced the Lagrangian description.

One disadvantage of this otherwise very successful formulation is that it makes a variational formulation of the water wave problem rather more difficult. Despite attempts going back to the 19th century (see the review by Salmon \cite{Salmon1988}), it took until the 1960s for Luke \cite{Luke1967} and Zakharov \cite{Zakharov1968} to formulate a Lagrangian and Hamiltonian for the water wave problem written in Eulerian coordinates. The Hamiltonian formulation was independently rediscovered by several authors, including Broer \cite{Broer1974} and Miles \cite{Miles1977}, the latter recovering the Hamiltonian from Luke's Lagrangian. Since these breakthroughs, variational principles in fluids have enjoyed a profusion of interest. 

Among the many remarkable contributions of Zakharov's 1968 paper \cite{Zakharov1968} is the derivation of a simplified (or reduced) Hamiltonian for waves of small steepness. This approach merged the long tradition of using perturbation theory (based on the notion of ``weak" nonlinearity) in water waves with the Hamiltonian framework, and led to the eponymous Zakharov equation, from which non-resonant contributions up to a given order have been suitably eliminated. This is effectively a compact way of writing the water wave problem in Fourier space up to a given order in nonlinearity, keeping only free waves (and so eliminating non-resonant, bound--wave components). 

The more classical perturbation approaches, such as that developed by Sir George Gabriel Stokes, proceed in physical space. In the simplest scenario, a single harmonic wave with argument $\xi = kx-\omega t,$ for $k$ the wavenumber, $\omega$ the frequency, and $x$ and $t$ space and time, is stipulated at the lowest order.  At higher orders in nonlinearity, multiples $2\xi$ and $3\xi$ of the fundamental harmonic appear as so-called \textit{bound harmonics}. It is possible in principle to introduce multiple harmonics $\xi_1 = k_1 x - \omega_1 t$ and $\xi_2 = k_2 x - \omega_2 t$ and so on at the leading order, and such an approach has been pursued by Tick \cite{Tick1959}, Longuet--Higgins \cite{Longuet-Higgins1962} and others, but quickly becomes cumbersome. In particular, bound modes of the form $2\xi_1, \, 2\xi_2, \, \xi_1+\xi_2, \, \xi_1-\xi_2$ will appear at second order, and still more at third order. 

The even more surprising fact is that, for waves on deep water, at the third order entirely new modes spontaneously appear! Even though you may reasonably start with only two modes, such as $\xi_1$ and $\xi_2$, if, say $2k_1 = k_2 + k_3$ for some new mode $k_3$ a process of energy transfer may cause the new mode to grow. This discovery of resonant wave-wave interaction by Phillips in 1960 \cite{Phillips1960} was a revolution in our understanding of surface water waves. As the details of this new resonant interaction theory began to be worked out, T.\ Brooke Benjamin and J.~E.\ Feir \cite{Benjamin1967a} discovered that this mechanism could explain difficulties in creating monochromatic waves of permanent form in the laboratory (see the review by Zakharov \& Ostrovsky \cite{Zakharov2009} for numerous interesting historical details).

This instability of the monochromatic (so-called Stokes) wave can be understood from the perspective of the nonlinear Schr\"odinger equation (NLS) for the envelope of deep-water waves. If initially small disturbances sufficiently close to a harmonic are generated, the envelope of the surface can be sensibly defined, and an equation derived which governs its behaviour to lowest order. Zakharov \cite{Zakharov1968} did just this, deriving NLS from a more general equation coming from the reduced Hamiltonian, and using a linear stability ansatz for NLS to investigate the Stokes wave simultaneously with Benjamin \& Feir.

More than 50 years on, Zakharov's equation is still being used in analytical explorations of water waves, and the Hamiltonian approach he pioneered has proved to be of major use in numerical wave modelling. Nevertheless, the Zakharov equation has a reputation for being difficult, with considerably more effort devoted to PDE daughter equations such as NLS, the Dysthe equation, and modifications thereof. The goal of the present review is to provide a gentle introduction to the Zakharov equation, with sufficient background for a reader with general knowledge of wave mechanics, and to show some of the results and techniques which have simple, practical value in describing deterministic waves in deep water. The Zakharov equation can also be profitably used for stochastic wave modelling, and applied in finite water depth, and the Discussion section takes up these important threads and gives pointers to the relevant literature. 

\subsection{The water wave problem}

We start with the full water wave problem for inviscid, incompressible flow, written in Eulerian variables:
\begin{subequations}
\begin{align}
&u_t + uu_x + vu_y + wu_z = -P_x\\
&v_t + uv_x + vv_y + wv_z = -P_y\\
&w_t + uw_x + vw_y + ww_z = -P_z - g\\
&u_x + v_y + w_z = 0 \\
&P = P_{0} \text{ on } z = \zeta(x,y,t) \\
&w = \zeta_t + u\zeta_x + v \zeta_y \text{ on } z = \zeta(x,y,t) \\
&w = 0 \text{ on } z = -h
\end{align}
\end{subequations}

The velocity field $\mathbf{u}=(u,v,w)$ in $(x,y,z)$-space is related to the pressure $P(x,y,z,t)$ and gravitational acceleration $g$ for a fluid of unit density $\rho =1$ kg/m$^3$. The fluid is said to occupy the domain between a flat bed at $z=-h$ and an as-yet unknown free-surface $\zeta(x,y,t)$, which moves in space and time. Above the surface of the water we assume a layer of air with constant pressure $P_0,$ which has no effect on the water below. These equations encompass the physically necessary assumptions that the water behaves according to the laws of Newtonian mechanics, that no fluid is created or destroyed, and that all velocities on the boundaries are purely tangential so that no fluid escapes the domain. This is -- with a few small changes -- the form of these equations which appears in the work of Leonhard Euler more than 250 years ago. In such generality it is difficult to progress, so we shall make a series of simplifying assumptions that will allow us to describe wave motion.

The first, very common simplifying assumption is the condition of irrotationality $\nabla \times \mathbf{u} = 0$ which implies the existence of a velocity potential $ \phi$ with $ \nabla \phi = \mathbf{u},$ reducing the field equation to Laplace's equation. Integrating the momentum equations with the help of the potential, this system can be rewritten as
\begin{subequations}
\begin{align} \label{eq:potential form eq 1}
&\Delta \phi = 0 \text{ on } -h < z < \zeta \\ \label{eq:potential form eq 2}
&\zeta_t + \nabla_x \phi \cdot \nabla_x \zeta = \phi_z \text{ on } z = \zeta \\ \label{eq:potential form eq 3}
&\phi_t + \frac{1}{2}(\nabla \phi)^2 + g\zeta = 0 \text{ on } z= \zeta \\ \label{eq:potential form eq 4}
&\phi_z = 0 \text{ on } z= -h
\end{align}
\end{subequations}
The horizontal ($x,y$) gradient is denoted $\nabla_x$ for clarity. In this step the pressure has been scaled out (by introducing a new pressure variable which measures deviations from hydrostatic pressure, and subsequently evaluating at the top boundary). The steps leading to this potential formulation are classical and can be found, for example, in Johnson \cite{Johnson1997}. The immediate advantage is that all nonlinearities have been moved to the boundary conditions. We will also note that, provided the water is deeper than about half a typical wavelength, the still-water depth $h$ can be taken to be infinite. This procedure, using
\begin{equation} \label{eq:potential form eq 4 DW}
\tag{\ref{eq:potential form eq 4}'} \phi_z \rightarrow 0 \text{ as } z \rightarrow \infty
\end{equation}
in place of \eqref{eq:potential form eq 4} simplifies calculations considerably, and we shall use it when convenient while commenting on its limitations.

If we want to find a Hamiltonian structure to the water wave problem in potential form, the natural place to look is the total energy. Indeed, expressions for kinetic and potential energy associated with wave motion can be found in classical textbooks such as Lamb's \textit{Hydrodynamics} \cite{Lamb1895}. Numerous attempts were made to find a variational formulation of the water wave problem (see the review by Miles \cite{Miles1981}), but it took until 1968 and the work of Zakharov \cite{Zakharov1968} to resolve the issue. This is a testament to the fact that the variational derivatives involved are somewhat tricky, and for this reason we present some of the steps of the calculation in the simplest setting.

\subsection{Hamiltonian formulation}
\label{ssec:Hamiltonian formulation}

The Hamiltonian is the rather simple expression
\begin{equation}
\label{eq:Hamiltonian}
H = T + V =  \int \int_{-h}^\zeta \frac{1}{2} |\nabla \phi|^2 dz + \frac{1}{2}g \zeta^2 dx, 
\end{equation}
which represents the total energy of the fluid. To simplify the algebra and unburden the notation we will drop the $y$-dependence, and work in a 2D domain $(x,z)$. Then the outermost integral over $x$ specifies which region of the fluid domain will be considered. This could be a finite region, say between $-L$ and $L$, in which case we would supply the problem with periodic lateral boundary conditions; it could equally well be an infinite region from $-\infty$ to $\infty$, in which case we should assume some decay, such as $\nabla \phi \rightarrow 0$ for $|x|\rightarrow \infty.$

The canonical variables will be defined, conveniently, as we shall see, at the free surface. They are $\zeta(x,t)$ and $\psi(x,t)=\phi(x,\zeta(x,t),t),$ and the canonical equations we expect are thus 
\begin{equation}
\frac{\partial \zeta}{\partial t} = \frac{\delta H}{\delta \psi}, \quad \frac{\partial \psi}{\partial t}=-\frac{\delta H}{\delta \zeta}.
\end{equation}
We shall show that these equations are equivalent to the surface boundary conditions \eqref{eq:potential form eq 2}, \eqref{eq:potential form eq 4}. The first problem seems to be that $H$ is written in terms of $\phi$ and not $\psi$, which makes the procedure to be pursued somewhat mysterious. There are elegant ways around this issue, chief among them the introduction of a Dirichlet-Neumann operator pursued by Craig \& Sulem \cite{Craig1993}, but for simplicity we follow the straightforward route employed by Broer \cite{Broer1974} which involves expressing variations $\delta \phi$ in terms of variations $\delta \psi.$

We write the variation of $H$ as 
\begin{equation}
\delta H = \int \int_{-h}^{\zeta} \phi_x (\delta \phi)_x + \phi_z (\delta \phi)_z dz dx +  \int \left[ \frac{1}{2} \left(\phi_x^2 + \phi_z^2\right)_{z=\zeta}  + g \zeta \right] \delta \zeta dx.
\end{equation}
In the first integral, the terms $\phi_z (\delta \phi)_z$ can immediately be treated by integration by parts, with the contribution on $z=-h$ vanishing due to \eqref{eq:potential form eq 4}. We note, using Leibniz' rule, that
\[ \frac{\partial}{\partial x} \int_{-h}^\zeta \phi_x (\delta \phi) dz = \int_{-h}^\zeta \phi_{xx} (\delta \phi) + \phi_x (\delta \phi)_x dz + \phi_x (\delta \phi)\vert_{z=\zeta} \zeta_x, \]
which allows us to reexpress the term $\phi_x (\delta \phi)_x.$ Using the Laplace equation \eqref{eq:potential form eq 1} eliminates the terms $\phi_{xx}$ and $\phi_{zz}$, while the lateral boundary condition means that 
\[   \left[ \int_{-h}^\zeta \phi_x (\delta \phi) dz \right]_{x=-L}^{x=L} = 0, \]
either due to periodicity if $L$ is finite, or due to the decay condition on $\nabla \phi$ if $L$ is taken to be $\infty.$ These simplifications yield
\begin{equation}
\delta H = \int \left[(\phi_z - \zeta_x \phi_x) \delta \phi\right]_{z = \zeta} dx +  \int \left[ \frac{1}{2} \left(\phi_x^2 + \phi_z^2\right)\delta \zeta + g\zeta  \delta \zeta \right]_{z=\zeta} dx.
\end{equation}
Since $\psi = \phi(x,\zeta,t),$ the above can be expressed in terms of variations $\delta \psi$ using $\delta \psi = \delta \phi\vert_{z=\zeta} + \phi_z \delta \zeta \vert_{z=\zeta}:$
\begin{equation}
\delta H = \int \left[(\phi_z - \zeta_x \phi_x) \delta \psi\right]_{z = \zeta} dx +  \int \left[ \frac{1}{2} \left( \left(\phi_x^2 + \phi_z^2\right) + g\zeta   - (\phi_z - \zeta_x \phi_x)\phi_z) \right) \delta \zeta \right]_{z=\zeta} dx.
\end{equation}
This last is a somewhat dubious manoeuvre, which can be avoided by  introduction of a Dirichlet-Neumann operator $G(\zeta)\psi = \phi_z - \zeta_x \phi_x\vert_{z=\zeta}$, see \cite{Craig1993}. (In this case the Hamiltonian \eqref{eq:Hamiltonian} is rewritten as $\frac{1}{2}\int \psi G(\zeta) \psi + g \zeta^2 dx,$ but we will not pursue this issue further, see \cite{Craig1993} for details.) 
From the variations $\delta \psi$ we immediately obtain 
\[ \frac{\delta H}{\delta \psi} = \phi_z - \zeta_x \phi_x = \zeta_t \text{ on } z = \zeta.\]
Likewise we find 
\[ \frac{\delta H}{\delta \zeta} = \left(\phi_x^2 + \phi_z^2\right) + g\zeta   - \underbrace{(\phi_z - \zeta_x \phi_x)}_{=\zeta_t}\phi_z = -\psi_t \text{ on } z = \zeta \]
by virtue of the fact that $\psi_t = \phi_t + \phi_z \zeta_t.$

The Hamiltonian is thus written in terms of the surface variables $\zeta$ and $\psi.$ To move from these to knowledge of the entire fluid velocity field we simply need to solve the boundary value problem for the Laplace equation:
\begin{align} \label{eq:phi-psi-coupling laplace}
&\phi_{xx} + \phi_{zz} = 0,\\ \label{eq:phi-psi-coupling sbc}
&\phi(x,z,t) = \psi(x,t) \text{ on } z = \zeta,\\ \label{eq:phi-psi-coupling bbc}
&\phi_z = 0 \text{ on } z = -h,
\end{align}
supplemented by either the decay or the periodicity condition at the lateral boundaries. This decoupling of the problem into a time-dependent part on the surface and a boundary value problem which can be solved at each time step is precisely the idea behind the very successful higher order spectral method, see \cite[Ch.\ 15]{Mei2005}. Analytical progress is also possible. 

Note that, if \eqref{eq:phi-psi-coupling sbc} were formulated on a flat surface $z=0$, as is the case for the linear system, it would be a simple matter to solve using a Fourier transform in $x$:
\[ \phi(x,z) = \frac{1}{2\pi} \int_{-\infty}^{\infty} e^{ikx} \hat{\psi}(k) \frac{\cosh(k(z+h))}{\cosh(kh)} dk, \]
with $\hat{\phantom{X}}$ denoting the Fourier transform. Alas, we have the boundary condition at the unknown surface $z = \zeta$ instead. Further analytical simplification now rests on making some assumptions. For example, if the wave slopes are assumed small, we could be tempted to rescale $\zeta \rightarrow \epsilon \zeta$ for a formal small parameter $\epsilon$ (later we shall see that the natural choice for $\epsilon$ is $ak$, the product of linear amplitude $a$ and wavenumber $k$) and subsequently Taylor expand the surface boundary condition around $z = 0:$
\[ \phi + \epsilon \zeta \phi_z + \frac{(\epsilon \zeta)^2}{2} \phi_{zz} + \ldots = \psi \text{ on } z = 0.\]
This is precisely the type of transfer of boundary conditions employed in the weakly nonlinear problem, and the problem could be tackled by solving iteratively using Fourier transforms.

Alternatively, the Hamiltonian can be formally expanded as set out by Krasitskii \cite{Krasitskii1994}. The first step in this procedure is to write the canonical equations in terms of the Fourier transformed variables $\hat{\zeta}(k)$ and $\hat{\phi}(k),$ as
\[ \frac{\partial \hat{\zeta}}{\partial t} = \frac{\delta H}{\delta \hat{\psi}^*}, \quad \frac{\partial \hat{\psi}}{\partial t} = - \frac{\delta H}{\delta \hat{\zeta}^*},\]
with $*$ denoting the complex conjugate. Then one defines a new pair of canonical variables $a(k)$ and $ia^*(k)$ via 
\[ \hat{\zeta} = \sqrt{\frac{\omega}{2g}} \left( a(k) + a^*(-k) \right), \quad \hat{\psi} = -i \sqrt{\frac{g}{2 \omega}} \left(a(k) - a^*(-k) \right), \]
where
\begin{equation} \label{eq: Linear Dispersion Relation}
\omega(\bk)^2 = g | \bk|
\end{equation}
is the linear dispersion relation for gravity waves in deep water, $\omega$ are the frequencies (in rad/s), and $g$ is the constant acceleration of gravity (taken to be 9.8 m/s$^2$ in computations).

The subsequent expansion of the Hamiltonian is quite lengthy (see Krasitskii \cite{Krasitskii1994}), but the principal idea is to transform from $a$ to a canonical variable $b$ which eliminates nonresonant terms, yielding a much less lengthy, reduced Hamiltonian, written in Fourier space up to a given order. The transformation from $a$ to $b$ in terms of an integral power series of general form can be found in \cite[p.\ 6 \& Sec.\ 3]{Krasitskii1994}. The cubic reduced Hamiltonian that emerges is 
\begin{equation} \label{eq:Krasitskii Hamiltonian}
H = \int \omega(k) b^*(k) b(k) dk + \frac{1}{2} \iiiint T(k,k_1,k_2,k_3) b^*(k) b^*(k_1) b(k_2) b(k_3) \delta(k+k_1-k_2-k_3) dk dk_1 dk_2 dk_3, 
\end{equation}
with the equation of motion
\begin{equation} \label{eq:ZE-continuous-autonomous}
i \frac{db(k)}{dt}=\frac{\delta H}{\delta b^*(k)}=\omega(k) b(k) + \iiint_{-\infty}^{\infty} T(k,k_1,k_2,k_3)b^*(k_1)b(k_2)b(k_3)\delta(k+k_1-k_2-k_3) dk_1 dk_2 dk_3.
\end{equation}
This is the cubic Zakharov equation (or reduced Zakharov equation), which is a remarkably compact way to encapsulate the water wave problem up to third order in Fourier space once all non-resonant contributions have been eliminated. In effect, all the heavy lifting in this otherwise generic equation is being done by the kernel term $T(k,k_1,k_2,k_3)$. Consequently, the kernel itself is very lengthy.
 Expressions for it were originally derived by Zakharov \cite{Zakharov1968}, albeit without all the symmetries necessary to preserve the Hamiltonian \eqref{eq:Krasitskii Hamiltonian}. Later derivations using perturbation methods, such as that by Yuen \& Lake \cite{Yuen1982} suffered from the same deficit, and the definitive Hamiltonian form of the kernels can be found in Krasitskii \cite{Krasitskii1994}.
 The wavenumber resonance condition $k+k_1=k_2+k_3$ is enforced by the Dirac delta-distribution $\delta(k+k_1-k_2-k_3),$ which is often abbreviated, using sub/superscripts to denote wavenumber dependence, as $\delta_{0,1}^{2,3}.$ 

We also note at this stage that the Zakharov equation is often written in the form 
\begin{equation} \label{eq:ZE-continuous-nonautonomous}
i \frac{dB_0}{dt}= \iiint_{-\infty}^{\infty} T_{0123} B_1^*B_2 B_3 e^{i\Delta_{0,1}^{2,3}t}\delta_{0,1}^{2,3} dk_1 dk_2 dk_3.
\end{equation}
Here we have introduced the compact notation $B_i = B(k_i,t),$ and more generally used subscripts to denote wavenumber components, such as abbreviating $T_{0123}=T(k_0,k_1,k_2,k_3)$. In addition, we denote by $\Delta_{0,1}^{2,3}=\omega_0 + \omega_1 - \omega_2 - \omega_3$ the frequency detuning. We can obtain the equation \eqref{eq:ZE-continuous-nonautonomous} from \eqref{eq:ZE-continuous-autonomous} via the transformation 
\[ b_i = B_i \exp(-i\omega_i t). \]

\subsection{Bound modes}
\label{ssec:Bound modes}

The reformulation of the water wave problem into the Zakharov equation \eqref{eq:ZE-continuous-autonomous} is relatively simple because it removes non-resonant contributions, such as bound modes.\footnote{This is accomplished in the transition between the variable $a$ to the variable $b$.} However, such bound modes are an essential part of a physically realistic solution, as they give rise to modifications of the wave shape. It is easily appreciated by any observer standing at the shore that waves on water -- even if these are very regular, as when observing swell from a distant storm -- look only superficially like sinusoids. In fact, the crests are sharper and the troughs are flatter, both effects that arise due to bound mode contributions. Thankfully the recovery of these bound modes can be accomplished with only a little calculation, which we demonstrate below (see Mei et al \cite[Ch.\ 14]{Mei2005} for a different approach using allied notation, as well as Janssen \cite{Janssen2009} for an account of the bound modes).

Recall from Section \ref{ssec:Hamiltonian formulation} that the complex amplitude $a(\bk,t)$ is related to the free surface $\zeta(\mathbf{x},t)$ by
\begin{equation} \label{eq:zeta_hat}
\hat{\zeta}({\bf k}, t) = \sqrt{\frac{\omega({\bf k})}{2g}} [a({\bf k}, t) + a^*(-{\bf k}, t)].
\end{equation}
For gravity water waves, we can write the complex amplitude $a$ in terms of 
\begin{equation} \label{eq:b expansion}
a({\bf k}, t) =e^{-i\omega ({ k})t} [ {B}({ k}) + 
B'({\bf k}) +  B^{\prime\prime}({ k})
+ \cdots]. 
\end{equation}
Here $B(\bk)$ is the variable appearing in the Zakharov equation \eqref{eq:ZE-continuous-nonautonomous}, from which non-resonant terms such as bound modes have been eliminated. The next term $B'(k)$ corresponds to all non-resonant second-order interactions (note that $'$ does not denote a derivative), i.e.\ the second order bound modes, and is an order smaller than $B(k),$ the wave steepness $\epsilon=a_0k_0$ being used as an ordering parameter, where $a_0$ is the linear wave amplitude and $k_0$ is the wavenumber. The next term $B''$ corresponds to all third-order bound modes, and is itself an order smaller than $B'$. We formally say $B=O(\epsilon), \, B'=O(\epsilon^2), \, B''=O(\epsilon^3),$ so that the lowest order free surface can be computed from $B$ alone (see Sections \ref{ssec:Stokes waves}--\ref{ssec:Bichromatic wave trains} for examples).

If the Zakharov equation \ref{eq:ZE-continuous-autonomous} can be solved for $B$, it is possible to recover the other contributions $B'$ and $B''$ from 
\begin{align}	\label{eq: B prime}
B' = -\iint_{-\infty}^\infty & \Bigg[V^{(1)}_{0,1,2} {B}_1 
{B}_2 \delta_{0-1-2} \frac{\exp[i(\omega - \omega_1 - 
\omega_2)t]}{\omega - \omega_1 - \omega_2}\nonumber\\
&+\, V^{(2)}_{0,1,2} {B}^*_1 {B}_2 \delta_{0+1-2} 
\frac{\exp[i(\omega + \omega_1 - \omega_2)t]}{\omega + \omega_1 - 
\omega_2}\nonumber\\
&+\, V^{(3)}_{0,1,2} {B}^*_1 {B}^*_2 \delta_{0+1+2} 
\frac{\exp[i(\omega + \omega_1 + \omega_2)t]}{\omega + \omega_1 + 
\omega_2}\Bigg] d{\bf k}_1\,d{\bf k}_2\,
\end{align}
and
\begin{align} \label{eq: B prime prime}
-iB^{\prime\prime} 
&= \iiint_{-\infty}^\infty \Big\{\tilde{T}^{(1)}_{0,1,2,3} 
{B}_1 {B}_2 \tilde{B}_3 \delta_{0-1-2-3} 
\frac{\exp[i(\omega - \omega_1 - \omega_2 - \omega_3)t]}{\omega - \omega_1 - \omega_2 - \omega_3}\nonumber\\ 
& +\, \Big(\tilde{T}^{(2)}_{0,1,2,3} - T_{0,1,2,3} \Big) {B}^*_1 
{B}_2 {B}_3 \delta_{0+1-2-3} \frac{\exp[i(\omega + \omega_1 - 
\omega_2 - \omega_3)t]}{\omega + \omega_1 - 
\omega_2 - \omega_3}\nonumber\\
& +\, \tilde{T}^{(3)}_{0,1,2,3} {B}^*_1 {B}^*_2 
{B}_3 \delta_{0+1+2-3} \frac{\exp[i(\omega + \omega_1 + \omega_2 - 
\omega_3)t]}{\omega + \omega_1 + \omega_2 - 
\omega_3}\nonumber\\
& +\, \tilde{T}^{(4)}_{0,1,2,3} {B}^*_1 {B}^*_2 
{B}^*_3 \delta_{0+1+2+3} \frac{\exp[i(\omega + \omega_1 + \omega_2 + 
\omega_3)t]}{\omega + \omega_1 + \omega_2 + 
\omega_3}\Big\} d{\bf k}_1 \,d{\bf k}_2 \,d{\bf k}_3. 
\end{align}
 The kernel $T_{0123}$ which appears is precisely that found in equation \eqref{eq:ZE-continuous-autonomous} and discussed in Appendix \ref{sec:Compact kernels}. The other kernels $V^{(1)}, \, V^{(2)}, \, V^{(3)}$ for nonresonant triad interactions and $\tilde T^{(1)}, \,\tilde T^{(2)}, \, \tilde T^{(3)}, \, \tilde T^{(4)}$ for nonresonant quartet interactions can be found in \cite[Appendix 14.B]{Mei2005}.
In this manner, the free surface elevation including bound waves can be recovered via
\begin{equation} \label{eq: zeta}
\zeta(x,t)=\frac{1}{2 \pi} \int_{-\infty}^{\infty} \sqrt{\frac{\omega(\bk)}{2g}} \left( \left( {B}(\bk) + B'(\bk) + B''(\bk) \right) e^{i(\bk \bx - \omega(\bk)t)} + \text{c.c} \right) d \bk. 
\end{equation}

\subsection{Discretising the Zakharov formulation}
\label{ssec:Discretisation}
For computational applications, it is imperative to have a discrete formulation of the Zakharov equation \eqref{eq:ZE-continuous-autonomous}. It turns out that this discrete formulation is also quite useful in the analysis of certain special cases. 

Our ansatz is
\begin{equation}
B(k,t) = \sum_{\xi = 1}^N B_\xi(t) \delta(k - k_\xi)
\end{equation}
with $\delta$ the Dirac delta distribution. This is sometimes referred to as the assumption of a discrete spectrum, where spectrum should be interpreted in the sense of a Fourier amplitude spectrum. Substituting into \eqref{eq:ZE-continuous-nonautonomous} yields
\begin{align} \nonumber 
i \sum_{\xi = 1}^N \frac{d B_\xi(t)}{dt} \delta(k - k_\xi) = & \iiint_{-\infty}^\infty T_{0123} \left( \sum_{\xi = 1}^N B^*_\xi(t) \delta(k_1 - k_\xi) \right)\left( \sum_{\xi = 1}^N B_\xi(t) \delta(k_2 - k_\xi) \right)\left( \sum_{\xi = 1}^N B_\xi(t) \delta(k_3 - k_\xi) \right) \\
& \cdot \delta(k + k_1 - k_2 -k_3) e^{i(\omega(k) + \omega(k_1) - \omega(k_2) - \omega(k_3))} dk_1 dk_2 dk_3,
\end{align}
which can be integrated once over wavenumber $k$ to yield the discretised Zakharov equation with the complex amplitudes $B_n = B(\bk_n,t)$:
\begin{equation} \label{eq: Discrete Zakharov Equation}
i \frac{dB_n}{dt} = \sum_{p,q,r =1}^N T_{npqr} \delta_{np}^{qr} e^{i \Delta_{npqr} t} B_p^* B_q B_r, \, n = 1, \, 2, \ldots , N,
\end{equation}
where $\delta_{np}^{qr}$ is now a Kronecker delta function such that 
\begin{equation*}
\delta_{np}^{qr} = \begin{cases} 1 \text{ for } \bk_n+\bk_p=\bk_q+\bk_r, \\
0 \text{ otherwise.} \end{cases}
\end{equation*}

The relation of the complex amplitudes $B_n$ to the free surface elevation $\eta$ is given in discrete form, to leading order (i.e.\ without bound contributions), by 
\begin{equation} \label{eq:Free Surface Transform}
\zeta = \frac{1}{2 \pi} \sum_n \left( \frac{|\bk_n|}{2 \omega_n} \right)^{1/2} \left( B_n e^{i(\bk_n \cdot x - \omega_n t)} + \text{c.c.} \right).
\end{equation}
The seemingly innocuous issue of discretising the Zakharov equation has been the object of some discussion, and the interested reader should consult \cite{Rasmussen1999} and \cite{Gramstad2011}.

\begin{figure}[ht!]
\centering
\includegraphics[width=0.9\linewidth]{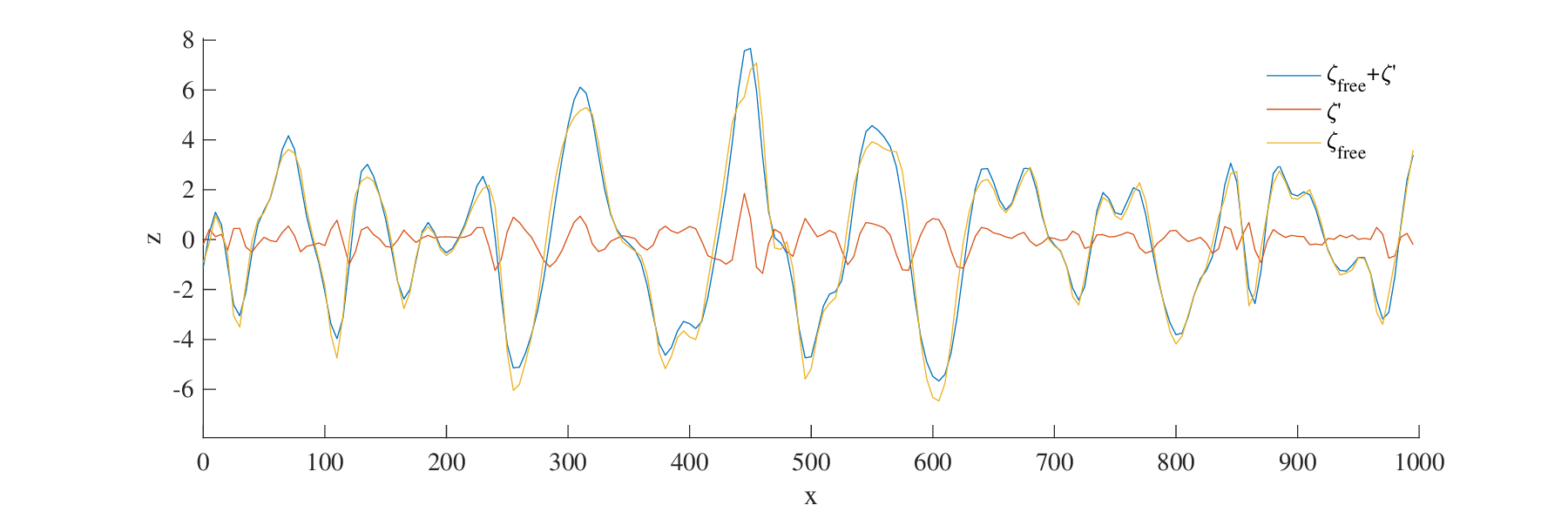}
\caption{Illustrative example showing a numerical computation of the free-wave surface elevation $\zeta_{free}$ together with the calculated second order bound waves $\zeta'$ and the total free surface $\zeta_{free}+\zeta'$, demonstrating how bound modes ``flatten the troughs" and ``sharpen the crests".}
\label{fig:Bound Modes Example}
\end{figure}

The expressions presented for the bound modes can be discretised in an analogous fashion, in short, replacing integrals by sums and Dirac's $\delta$ by Kronecker's $\delta$. This leads essentially to expressions for the full second or third-order free surface, akin to those presented by Dalzell \cite{Dalzell1999} for the determinstic scenario or Janssen \cite{Janssen2009} for the case of energy spectra, both to second order. These expressions can be quite useful for numerical calculation of bound modes, for example when a simulation of a realistic sea state is desired (see Figure \ref{fig:Bound Modes Example}).

\section{Simple solutions of the Zakharov equation}
\label{sec:Simple solutions}

\subsection{Stokes' third order wave in infinite depth}
\label{ssec:Stokes waves}
In order to acquire some feeling for how the Zakharov formulation works it is useful to see how well-known results can be recovered from it. The simplest such scenario consists in assuming a single wavenumber and frequency (a discretisation with one mode, see Section \ref{ssec:Discretisation}), and thus recovering the free and bound parts of the classical Stokes' expansion for deep water waves. 

When viewed from the perspective of classical perturbation theory as developed by Stokes and used extensively in fluid mechanics over the course of the last 150 years it is somewhat counterintuitive that Stokes waves would have anything to do with wave interaction.
This is because the viewpoint of perturbation theory makes an \textit{a priori} assumption about which Fourier modes are involved, and so precludes any interaction except between a Fourier mode and itself. That this ``self-interactin" is a meaningful interaction at all only became clear with work of Tick \cite{Tick1959}, Phillips \cite{Phillips1960}, and others starting in the late 1950s. Indeed, the resonant interaction theory stipulates that $\bk_1 + \bk_2 = \bk_3 + \bk_4$ and $\omega_1 + \omega_2 = \omega_3 + \omega_4$ (with the latter being fulfilled up to $O(\epsilon^2)$ in so-called near resonance), which is trivially the case if $\bk_1=\bk_2=\bk_3=\bk_4.$

To this end, we will assume a single wave mode only, such that 
\begin{equation} \label{eq: Stokes ansatz}
{B}(\bk,t)=b(t) \delta(\bk-\bk_0).
\end{equation}
Inserting into \eqref{eq:ZE-continuous-nonautonomous} one finds the solution
\begin{equation} \label{eq: Stokes wave - 1st order}
 b = B_0 e^{-i T_{0000} |B_0|^2 t},
\end{equation}
where $B_0 = b(k_0,t=0)$ is the initial value of the complex amplitude.
Inserting the same ansatz \eqref{eq: Stokes ansatz} into \eqref{eq: B prime} yields
\begin{align} \nonumber 
-B' & =  V^{(1)}(k,k_0,k_0) b^2 \delta(k-2k_0) \frac{e^{i(\omega(k) - 2\omega(k_0))t}}{\omega(k) - 2\omega(k_0)} + V^{(2)}(k,k_0,k_0) |b|^2 \delta(k) \frac{e^{i \omega(k) t}}{\omega(k)} \\ \label{eq: Stokes wave - 2nd order}
& + V^{(3)}(k,k_0,k_0) (b^*)^2 \delta(k+2k_0) \frac{e^{i(\omega(k) + 2\omega(k_0))t}}{\omega(k) + 2\omega(k_0)}
\end{align}
Finally, inserting ansatz \eqref{eq: Stokes ansatz} into \eqref{eq: B prime prime} yields
\begin{align}\nonumber
-B'' &= \tilde{T}^{(1)}(k,k_0,k_0,k_0) b^3 \delta(k-3k_0) \frac{e^{i(\omega(k) - 3 \omega(k_0))t}}{\omega(k) - 3 \omega(k_0)} \\\nonumber
&+ \left(\tilde{T}^{(2)}(k,k_0,k_0,k_0)-T(k,k_0,k_0,k_0)\right) |b|^2 b \delta(k-k_0) \frac{e^{i(\omega(k) - \omega(k_0))t}}{\omega(k) - \omega(k_0)} \\\nonumber
&+ \tilde{T}^{(3)}(k,k_0,k_0,k_0) |b|^2 b^* \delta(k+k_0) \frac{e^{i(\omega(k) + \omega(k_0))t}}{\omega(k) +  \omega(k_0)} \\
&+ \tilde{T}^{(4)}(k,k_0,k_0,k_0) (b^*)^3 \delta(k+3k_0) \frac{e^{i(\omega(k) + 3 \omega(k_0))t}}{\omega(k) + 3 \omega(k_0)} \label{eq: Stokes wave - 3rd order}
\end{align}
Inserting these into \eqref{eq: zeta} then yields expressions for the free surface, which we decompose  into 
\[ \zeta = \zeta_{\text{free}} + \zeta' + \zeta'' + \ldots \]
Plugging \eqref{eq: Stokes wave - 1st order} into \eqref{eq: zeta} we find the free surface without bound modes
\begin{equation}
\zeta_{\text{free}}(x,t) = \frac{B_0}{\pi} \sqrt{\frac{\omega_0}{2 g}} \cos \left(k_0 x - \omega_0 t - T_{0000} |B_0|^2 t \right). 
\end{equation}
This is simply a sinusoid with a frequency correction -- reflecting the well-known fact that nonlinear waves have a celerity which depends on their amplitude. It is natural to define
\begin{equation} \label{eq:B-to-a}
\frac{B_0}{\pi} \sqrt{\frac{\omega_0}{2g}} = a_0
\end{equation}
and regard $a_0$ as the amplitude of the wave.

In a similar vein, plugging \eqref{eq: Stokes wave - 2nd order} into \eqref{eq: zeta}, resolving the $\delta$ functions, replacing $b$ by the solution \eqref{eq: Stokes wave - 1st order}, and using the fact that 
\[ V^{(1)}(2k,k,k) = V^{(3)}(-2k,k,k) = \frac{1}{8 \pi} \left( 8 g \|k\|^7\right)^{1/4}\]
we find
\begin{equation}
\zeta'(x,t) = \frac{1}{8 \pi^2} \sqrt{\frac{\omega_2}{2g}} \frac{\omega_2}{g k_0} (8g |k_0|^7)^{1/4} B_0^2 \cos (2k_0 x - 2 \omega_0 t - 2 T_{0000}|B_0|^2 t),
\end{equation}
where $\omega_2 = \omega(2k_0), \, \omega_0 = \omega(k_0)$ are used as convenient abbreviations.

Repeating the same procedure for \eqref{eq: Stokes wave - 3rd order}, and using the fact that 
\[ \tilde{T}^{(1)}(3k,k,k,k) = \tilde{T}^{(4)}(-3k,k,k,k) = \frac{3^{5/4}\|k\|^3}{16 \pi^2}, \quad \tilde{T}^{(3)}(-k,k,k,k) = -\frac{ \|k\|^3}{8 \pi^2},\]
we find 
\begin{align*}
\zeta''(x,t)= \frac{B_0^3}{\pi} \left[ \sqrt{\frac{\omega_3}{2 g}} \frac{3^{5/4} |k_0|^3}{16 \pi^2} \frac{ \omega_3}{3 g k_0}   \cos(3(k_0 x - \omega_0 t - T_{0000} |B_0|^2 t)) \right. \\ \left.
+ \sqrt{\frac{\omega_0}{2g}} \frac{|k_0|^3}{16 \pi^2 \omega_0}   \cos(k_0 x - \omega_0 t - T_{0,0,0,0} |B_0|^2 t) \right].
\end{align*}
Here we abbreviate $\omega_3=\omega(3k_0).$

\begin{figure}
\centering
\includegraphics[width=0.8\linewidth]{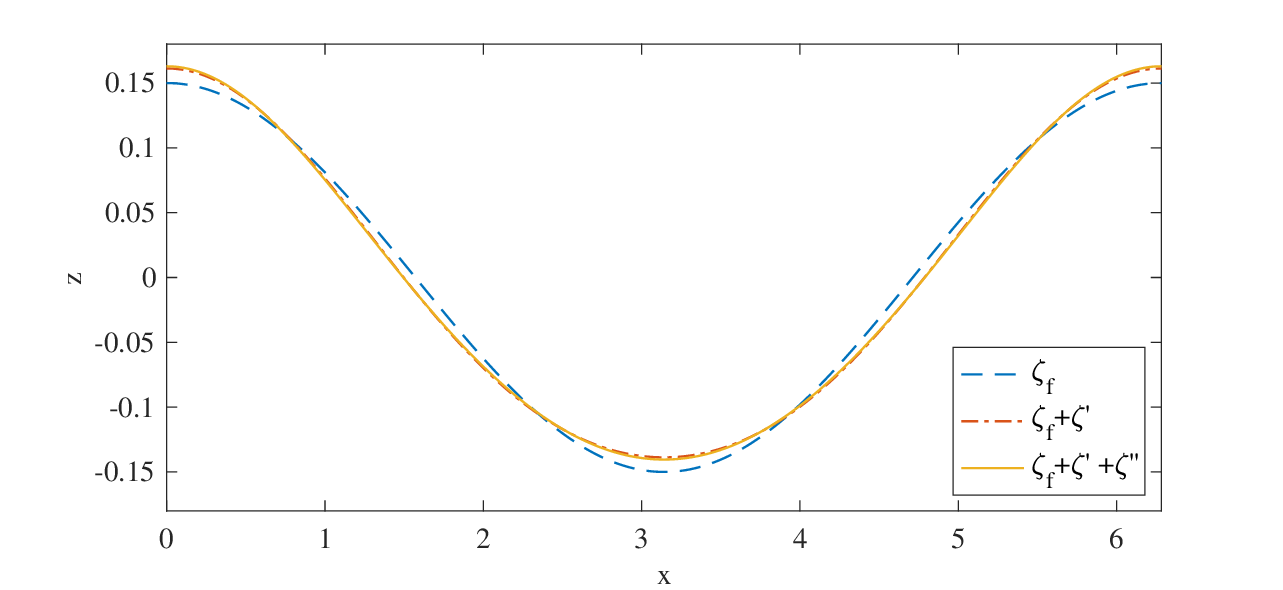}
\caption{Monochromatic wave train shown with free modes only ($\zeta_f$), with second order bound modes ($\zeta_f+\zeta'$), and with second and third order bound modes ($\zeta_f+\zeta'+\zeta''$). The wavenumber $k_0 = 1 \, \text{m}^{-1}$ and the linear amplitude $a_0 = 0.15$ m.}
\label{fig:Stokes bound modes}
\end{figure}

The expressions for the bound components up to third order are thus: 
\begin{align} \label{eq:Stokes wave zeta free}
&\zeta_{\text{free}}= a_0 \cos \left(k_0 x - \omega_0 t - T_{0000} |B_0|^2 t \right), \\ \label{eq:Stokes wave zeta prime}
&\zeta' = \frac{a_0^2 |k_0|}{2} \cos (2k_0 x - 2 \omega_0 t - 2 T_{0000}|B_0|^2 t), \\ \label{eq:Stokes wave zeta prime prime}
&\zeta'' = \frac{3 |k_0|^2 a_0^3}{8} \cos(3(k_0 x - \omega_0 t - T_{0000} |B_0|^2 t)) + \frac{|k_0|^2 a_0^3}{8} \cos(k_0 x - \omega_0 t - T_{0000} |B_0|^2 t).
\end{align}
Finally the frequency correction should be simplified by using the identity
\[ T_{0000} = \frac{|k_0|^3}{4 \pi^2},\]
which agrees with the Stokes' wave solution as found in Wehausen \& Laitone \cite[Eq.\ (27.25)]{Wehausen1960}. Note that the second-order bound modes modify the (linear) free wave solution by an $O(\epsilon)$ correction, while the third order bound modes modify it by an even smaller $O(\epsilon^2)$ correction, as shown in Figure \ref{fig:Stokes bound modes}.

It is worth noting that the fourth order terms included in the Zakharov formulation by Krasitskii \cite{Krasitskii1994} will not change the frequency corrections (and thus $\zeta_{free}$) in the Stokes' solution given above. The simple reason is that a single wave cannot interact with itself at fourth order, since $\bk_1 + \bk_1 \neq \bk_1 + \bk_1 + \bk_1.$ Further frequency corrections will appear at fifth order, where the interactions are between sextets of waves, and where a Zakharov formulation should recover the corrections found, e.g.\ by Fenton \cite{Fenton1985}.

\subsection{Bichromatic wave trains}
\label{ssec:Bichromatic wave trains}

After the single mode, the next step in complexity is to assume that two Fourier modes $k_1$ and $k_2$ are initially present. For the linear water wave problem this is trivial, of course, but the expansion to higher order can become quite cumbersome, as shown in the work of Longuet-Higgins \& Phillips \cite{Longuet-Higgins1962d} (with algebraic details given in Longuet-Higgins \cite{Longuet-Higgins1962}).

In comparison we shall see that the Zakharov formulation makes light work of this case. We consider the interaction of two weakly nonlinear wavetrains,  Making the discrete spectrum ansatz
\begin{equation}
\label{eq_14.6.1}
B(k,t) = B_1(t) \delta({ k} - { k}_1) + B_2(t) 
\delta({ k} - { k}_2) 
\end{equation}
and substituting \eqref{eq_14.6.1} into \eqref{eq:ZE-continuous-nonautonomous}, we find
\begin{align}	
\label{eq_14.6.2}
i\frac{dB_1}{dt} &= T_{1111} |B_1|^2 B_1 + 2 T_{1212}  |B_2|^2 B_1,\\
\label{eq_14.6.3}
i\frac{dB_2}{dt} &= T_{2222} |B_2|^2 B_2 + 2T_{1212} |B_1|^2 B_2. 
\end{align}
In what follows, we denote $T_{1111}$ by $T_1,$ $T_{2222}$ by $T_2,$ and $T_{1212}$ by $T_{12},$ and note that by symmetry of the kernel $T_{12}=T_{21}$ (see Appendix \ref{sec:Compact kernels}). Although the system of ordinary differential equations (\ref{eq_14.6.2}) and (\ref{eq_14.6.3}) is coupled and nonlinear, it (perhaps somewhat surprisingly) nevertheless has a solution with constant amplitudes given by
\begin{align}	
\label{eq_14.6.4}
B_1(t) &= B_1(0) \exp(-i(T_1 B_1(0)^2 + 2T_{12} B_2(0)^2)t),\\ 
\label{eq_14.6.5}	
B_2(t) &= B_2(0) \exp(-i(T_2 B_1(0)^2 + 2T_{12} B_2(0)^2)t). 
\end{align}
We now substitute \eqref{eq_14.6.1}, \eqref{eq_14.6.4} and \eqref{eq_14.6.5} in \eqref{eq:Free Surface Transform} and write the resulting surface elevation without bound modes as 
\begin{equation}	
\label{eq_14.6.6}
\zeta_{\text{free}}(x, t) = a_1 \cos(k_1 x - \Omega_1 t) + a_2 \cos(k_2 x- \Omega_2 t)
\end{equation}
where $a_1$ and $a_2$ represent the amplitudes of the two wavetrains via \eqref{eq:B-to-a}:
\begin{align}\label{eq_14.6.7}
B_1(0) &=  2\pi\left(\frac{\omega_a}{2k_a}\right)^{1/2} 
a_1,\\
B_2(0) &=  2\pi\left(\frac{\omega_b}{2k_b}\right)^{1/2} a_2.
\end{align}

The nonlinear corrected frequencies are given by 
\begin{align}
\label{eq_14.6.8}
\Omega_1 &= \omega_1 + T_1 B_1(0)^2 + 2T_{12} B_2(0)^2,\\
\label{eq_14.6.9}
\Omega_2 &= \omega_2 + T_2 B_2(0)^2 + 2T_{12} B_1(0)^2.
\end{align}
The first term is simply the linear frequency, and the second term in each expression is due to the self-self interaction which gives rise to the Stokes wave discussed in Section \ref{ssec:Stokes waves}. The third term, which involves the kernel $T_{12}$ is a mutual interaction between waves $k_1$ and $k_2.$ This result, first found by Longuet-Higgins and Phillips \cite{Longuet-Higgins1962}, demonstrates how one train of waves affects the frequency of another, at exactly the same order as the well-known Stokes' correction. 

For unidirectional waves we can use the particularly simple form of the two-wave interaction kernel $T_{12}$ (see \eqref{eq:Kernel-unidir-two-wave}) to write (for $\epsilon_i = a_i k_i$) 
\begin{align}\label{eq:fixed-Omega1}
&\Omega_1 = \omega_1\left(1 + \frac{\epsilon_1^2}{2} + \frac{\omega_2}{\omega_1}\cdot \frac{k_1}{k_2}\cdot \epsilon_2^2\right),\\
&\Omega_2 = \omega_2\left(1 + \frac{\epsilon_2^2}{2} + \frac{\omega_1}{\omega_2}\cdot \frac{k_2^2}{k_1^2}\cdot \epsilon_1^2\right)  \label{eq:fixed-Omega2}
\end{align}
where $k_1 > k_2$. Inspection of these expressions shows that the longer wave $k_2$ has a greater effect on the frequency of the shorter wave $k_1$ than vice versa -- there is an in-built asymmetry in how the waves affect one another. We shall see in Section \ref{sec:Dispersion corrections} how this idea can be extended to configurations with multiple waves, including a continuous spectrum of waves.

\begin{figure}
\centering
\includegraphics[width=0.7\linewidth]{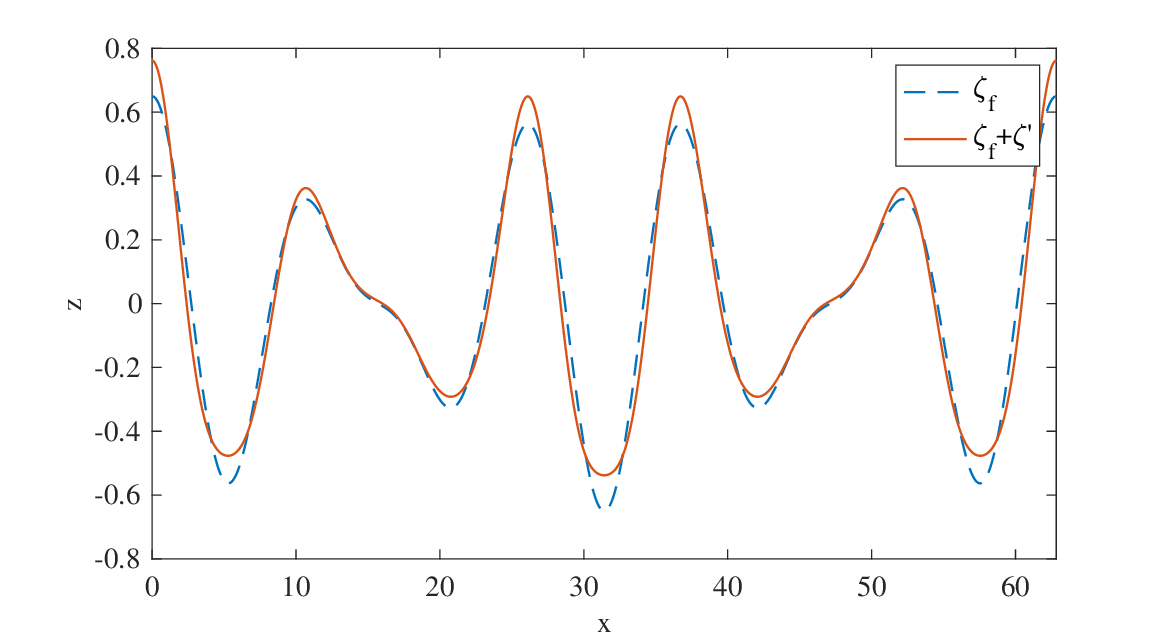}
\caption{Bichromatic sea-state with free modes only ($\zeta_f$) and second order bound modes computed with \eqref{eq:Bichromatic-2nd order free surface}.  The wavenumbers are $k_1 = 0.7 \text{ m}^{-1}, \, k_2 = 0.5 \text{ m}^{-1}$, the linear amplitudes are $a_1 = 0.25$ m, $a_2 = 0.4$ m (high steepness chosen for emphasis).}
\label{fig:Bichrom-bound-modes}
\end{figure}

One can also plug the discretisation \eqref{eq_14.6.1} into the expression for the second order contributions \eqref{eq: B prime}, resolve the delta-distributions, and insert the resulting expression for $B'(k,t)$ into equation \eqref{eq: zeta} to compute the second order free surface. Using the kernel identities for scalar wavenumbers
\begin{align*}
& V^{(1)}_{2a,a,a} = \frac{8^{3/4} k_a^{7/4} g^{1/4}}{\pi}, \quad V^{(1)}_{a+b,a,b} = V^{(1)}_{a+b,b,a} = \frac{\sqrt{2}((k_a+k_b)k_a^3 k_b^3 g)^{1/4}}{8 \pi},\\
& V^{(2)}_{a-b,b,a} = -V^{(2)}_{b-a,a,b} = \frac{\sqrt{2}((k_a-k_b)^3 k_b^3 k_a g)^{1/4}}{4 \pi}, \\
& V^{(3)}_{-2a,a,a} = \frac{8^{3/4} k_a^{7/4} g^{1/4}}{\pi}, \quad V^{(3)}_{-a-b,a,b} = \frac{\sqrt{2}((k_a+k_b)k_a^3k_b^3 g)^{1/4}}{8 \pi},
\end{align*}
(without loss of generality $k_a > k_b$) it is possible after some algebra to obtain the expression for the second-order bound modes
\begin{align} \label{eq:Bichromatic-2nd order free surface}
\zeta'(x,t) = \frac{a_a^2 k_a}{2} \cos(2\Xi_a) + \frac{a_b^2 k_b}{2} \cos(2 \Xi_b) + \frac{a_a a_b}{2} \left( (k_a + k_b) \cos(\Xi_a + \Xi_b) - (k_a - k_b) \cos(\Xi_a - \Xi_b)\right).
\end{align}
Here $\Xi_i = k_i x - \Omega_i t,$ with $\Omega_i$ the respective nonlinear corrected frequency. The first two terms in \eqref{eq:Bichromatic-2nd order free surface} are clearly the analogues of the single-mode superharmonics appearing in the Stokes wave (see \eqref{eq:Stokes wave zeta prime}), to which are added the superharmonic term (with $k_a + k_b$) and the subharmonic term (with $k_a - k_b$), the latter being associated with a set-down (note the minus term). A plot of these contributions can be seen in Figure \ref{fig:Bichrom-bound-modes}.

\section{Discrete wave-wave interactions}
\label{sec:Discrete interactions}

We have discussed how to discretise the Zakharov equation in Section \ref{ssec:Discretisation}, and seen some simple examples where explicit solutions of the Zakharov equation can be found in Section \ref{sec:Simple solutions}. However, despite our contention that the Zakharov equation describes wave interactions the systems described hitherto showcase only the most basic interaction: a correction to the frequency due to the effects of finite but small amplitude. As was suggested in the Introduction, additional modes will change this picture, and allow for nontrivial resonances beyond $k_1 + k_1 = k_1 + k_1$ and $k_1 + k_2 = k_1 + k_2.$

\subsection{Benjamin-Feir instability}
\label{ssec:BFI}

One of the most famous applications of discrete wave interactions is in the description of the instability of monochromatic waves on the surface of deep water. 
Taking three equally spaced modes initially, $k_0,\, k_{\pm 1},$ such that $\bk_0 + \bk_0 = \bk_{+1} + \bk_{-1},$ the full Hamiltonian, or equivalently the full system of three ODEs, is somewhat lengthy. We initially retain only the principal mode $k_0$, assuming $b_0 \gg b_{\pm1}$, which yields from \eqref{eq:ZE-continuous-autonomous}
\begin{equation}
\label{eq:Lin 5 Wave-b0}
i \frac{d b_0}{dt} = \omega_0 b_0 + T_0 b_0 |b_0|^2,
\end{equation}
the well known equation for the Stokes' wave found in Section \ref{ssec:Stokes waves}.

To understand the initial influence of small disturbances, we perform a linear stability analysis about this solution \ref{eq:Lin 5 Wave-b0}. This means retaining in the system of ODEs only those terms linear in $b_{\pm 1}.$ This yields the system
\begin{align} \label{eq:2nd O 5 Wave b1}
i \frac{db_1}{dt} &= \left( \omega_1 + 2 T_{0,1} |b_0|^2 \right) b_1 + T_{-1,1,0,0} b_{-1}^* b_0^2 \\ \label{eq:2nd O 5 Wave b-1}
i \frac{db_{-1}}{dt} &= \left( \omega_{-1} + 2 T_{0,-1} |b_0|^2 \right) b_{-1} + T_{-1,1,0,0} b_{1}^* b_0^2 
\end{align}

The solution to the lowest order equation \eqref{eq:Lin 5 Wave-b0} is the Stokes' wave:
\begin{equation}
b_0 = \beta_0 \exp(-i\Omega_0 t)
\end{equation}
where we use the abbreviations
\begin{align*}
\Omega_0 &= \omega_0 + T_0 |\beta_0|^2,\\
\Omega_1 &= \omega_1 + T_{0,1} |\beta_0|^2,\\
\Omega_{-1} &= \omega_{-1} + T_{0,-1} |\beta_0|^2.
\end{align*}
\begin{figure}
\centering
\includegraphics[width=0.45\linewidth]{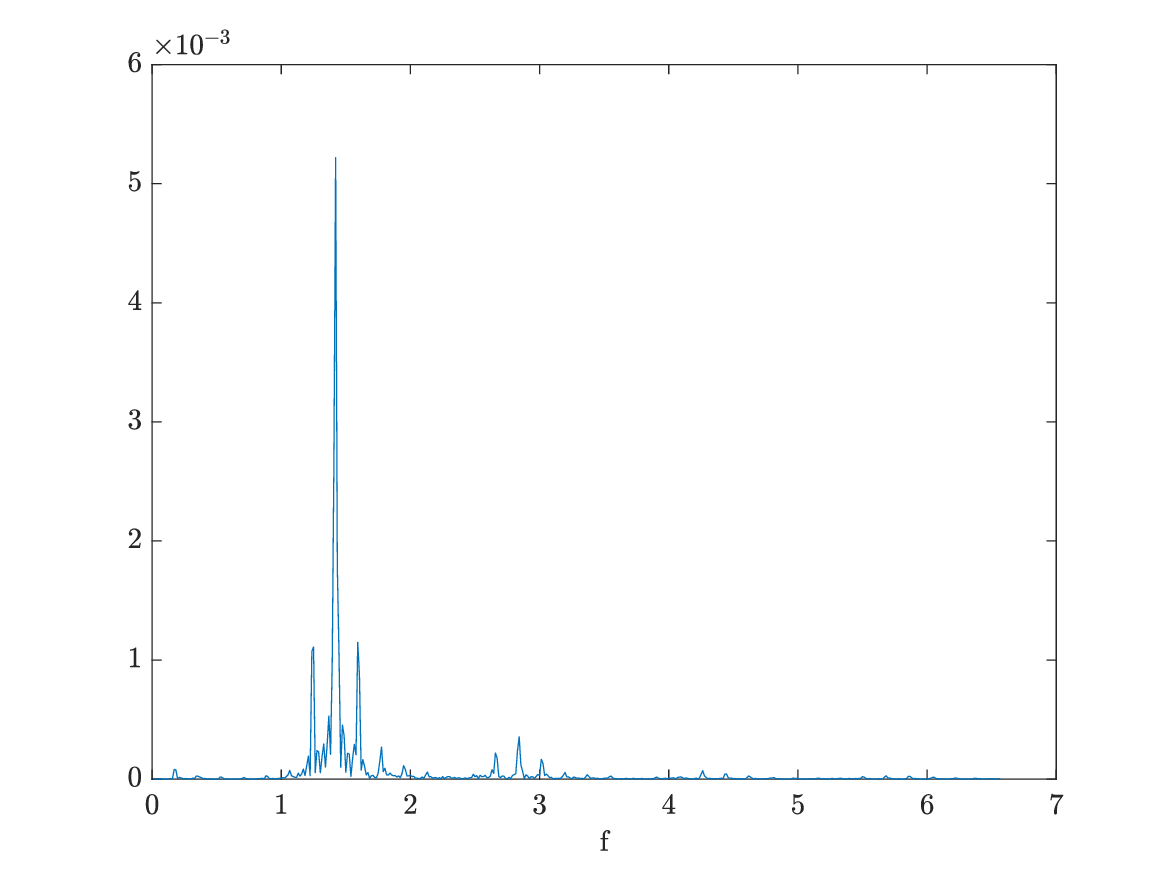}
\includegraphics[width=0.45\linewidth]{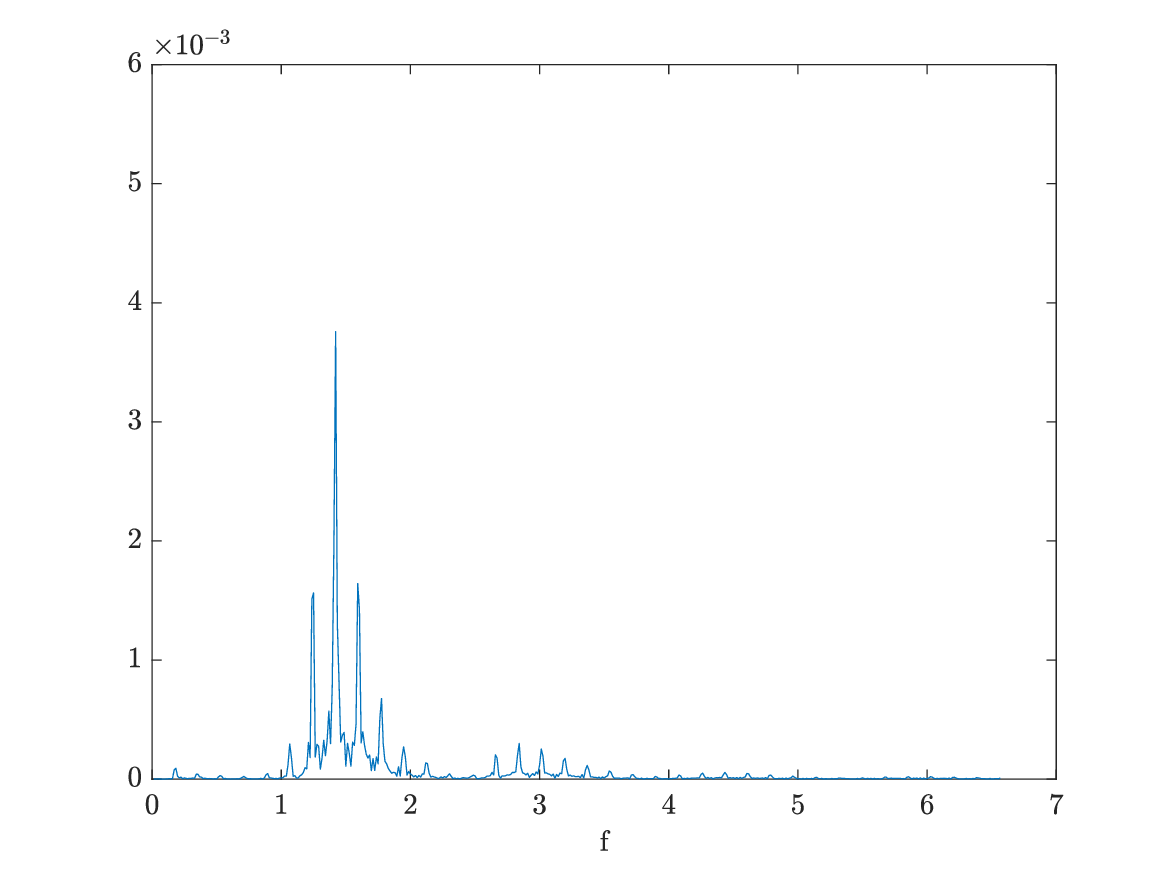}
\caption{Fourier amplitude spectrum of a monochromatic (Stokes) wave perturbed by two small side bands from wave-flume measurements at a probe $x=0$ m (left panel) and a probe $x=5$ m (right panel), showing the distinct growth of the side-bands with propagation distance due to the Benjamin-Feir instability .}
\label{fig:BFI}
\end{figure}
We rewrite the system \eqref{eq:2nd O 5 Wave b1}--\eqref{eq:2nd O 5 Wave b-1} as 
\begin{align} 
i \frac{db_1}{dt} &= \Omega_1 b_1 + T_{-1,1,0,0} b_{-1}^* b_0^2 \\ 
i \frac{db_{-1}}{dt} &= \Omega_{-1}
 b_{-1} + T_{-1,1,0,0} b_{1}^* b_0^2 
\end{align}
and insert the ansatz
\begin{align*}
b_1(t) &= \beta_1 e^{\sigma t} e^{-i \Omega_1 t} e^{-i \frac{\Delta}{2} t} \\
b_{-1}(t) &= \beta_{-1} e^{\sigma^* t} e^{-i \Omega_{-1} t} e^{-i \frac{\Delta}{2} t} \\
\end{align*}
with $\Delta = 2\Omega_0 - \Omega_1 - \Omega_{-1}.$ This yields the following linear system 
\begin{align*}
\beta_1 (i\sigma + \frac{\Delta}{2} ) - T_{-1,1,0,0} \beta_{-1}^* \beta_0^2 = 0,\\
\beta_{-1} (i \sigma^* + \frac{\Delta}{2}) - T_{-1,1,0,0} \beta_1^* \beta_0^2=0.
\end{align*}
This has a nontrivial solution only if the determinant of the coefficient matrix 
\[ \begin{pmatrix}
i \sigma + \frac{\Delta}{2} & -T_{-1,1,0,0} \beta_0^2 \\
-T_{-1,1,0,0} (\beta_0^*)^2 & -i\sigma + \frac{\Delta}{2} 
\end{pmatrix}
\begin{pmatrix}
\beta_1 \\ \beta_{-1}^* 
\end{pmatrix}
= \begin{pmatrix}
0\\0
\end{pmatrix}
\]
vanishes. This is equivalent to 
\begin{equation}
\sigma^2 = T^2_{-1,1,0,0} |\beta_0|^4 - \frac{\Delta^2}{4},
\end{equation}
which determines the instability growth rate. If the right-hand side is positive we have instability (a positive growth rate $\sigma$), while a negative right-hand side yields stability (imaginary $\sigma$ and so oscillatory solutions). Note that this also determines a relationship between $\beta_1$ and $\beta_{-1},$ and ensures that these modes grow at the same rate. This result about the instability of the Stokes wave was first found via perturbation theory and subsequent experiments by Benjamin \& Feir \cite{Benjamin1967a}, and Figure \ref{fig:BFI} shows Fourier amplitude spectra from a similar flume experiment.\footnote{The attentive reader may object that the evolution in the flume is in space, whereas our equations are in time; see the Discussion or Shemer \cite{Shemer2016}.} The instability criterion for a narrow-banded case (starting from the NLS) was given by Zakharov \cite{Zakharov1968}, and the present, more general result, is due to Yuen \& Lake \cite{Yuen1982}. 

This classical result is, of course, only the beginning of the energy exchange among a particular (so-called degenerate) quartet of waves $\bk_0 + \bk_0 = \bk_{+1} + \bk_{-1}$. The initial growth is exponential, and the assumptions of smallness inherent in the linearisation are soon violated. It is natural to ask what happens thereafter, and attempts have been made to answer this question via numerical simulation since the 1970s (see particularly the review by Yuen \& Lake \cite{Yuen1982}). In the next section we will take a somewhat different approach, and show how to tackle this problem analytically via a simple reformulation.

\subsection{Action-angle reformulation and reduction to a planar system}
\label{ssec:Action-angle reduction}

Some of the first studies of nonlinear resonant interactions focused on instabilities, such as that discussed in Section \ref{ssec:BFI} above, usually in terms of linear stability analysis. By definition, an instability must start somewhere, that is it must start with a solution which can then be perturbed. The solutions available to us so far are essentially the one and two-mode cases discussed in Sections \ref{ssec:Stokes waves} and \ref{ssec:Bichromatic wave trains}, and we have already seen how the former plays a role in the Benjamin-Feir instability. In an elegant paper Leblanc \cite{Leblanc2009} also showed how the bichromatic wave trains can be treated by means of linear stability analysis of the Zakharov equation, giving rise to so-called Type Ia and Type Ib instabilities.

In all these treatments of instability the question remains what happens to the unstable solution after the initial exponential growth. Remarkably, the answer can be found explicitly by a clever reduction of the system, which exploits the conservation laws. Such reduction of a system of resonantly interacting waves goes back at least to Bretherton \cite{Bretherton1964a}, and was first employed on a discrete modified (quartic) Zakharov equation in an exploration of Type II instability \cite{Shemer1985}.  Some 20 years on, Stiassnie \& Shemer \cite{Stiassnie2005} used this approach on the cubic Zakharov equation to understand the interactions of four waves. Our focus will be on cubically nonlinear interactions only.

The starting point of the reduction is to consider the conserved quantities which enable progress on this problem. The first and arguably most important is the Hamiltonian (total energy) written in discrete formulation as
\begin{align} \label{eq:Hamiltonian discrete:original}
H(b_1\ldots b_N,b^*_1\ldots b^*_N) = \sum_{i=1}^N \omega_i |b_i|^2 + \frac{1}{2}\sum_{i,j,k,l=1}^N T_{ijkl} \delta_{ij}^{kl} b_i^* b_j^* b_k b_l.
\end{align}

Note that the discrete Zakharov equation can be obtained directly from this Hamiltonian as
\begin{align}
    i\frac{db_i}{dt} = \frac{\partial H}{\partial b_i^*} = \omega_i b_i + \sum_{j,k,l=1}^N T_{ijkl} \delta_{ij}^{kl} b_j^* b_k b_l,\quad\text{for $i = 1,\ldots,N$.}
\end{align}

In addition to the Hamiltonian, the momentum $\mathbf{M}$ and total wave action $A$ are conserved (see \cite[Eq.\ (3.36)ff]{Krasitskii1994}) when cubically nonlinear interactions only are taken into account:
\begin{align}\label{eq:invariants}
    \mathbf{M} =\sum_{i=1}^4 \bk_i |b_i|^2, && A= \sum_{i=1}^4  |b_i|^2.
\end{align}

The Zakharov Hamiltonian in action-angle variables $b_i = |b_i|\exp(-i\phi_i)$ is 
\begin{align}\label{eq:Hamiltonian discrete}
\begin{aligned}    
H(|b_1|^2,\ldots,|b_N|^2,\phi_1,\ldots,\phi_N) &= \sum_{i=1}^N \omega_i|b_i|^2 + \frac{1}{2}\sum_{i,j=1}^N e_{ij}T_{ij}|b_i|^2|b_j|^2 + \\
    &+ \frac{1}{2}\sum_{i,j=1}^N \sum_{k\neq i,j} \sum_{l\neq i,j} T_{ijkl} \sqrt{|b_i|^2|b_j|^2|b_k|^2|b_l|^2}  \delta_{ij}^{kl} \cos(\phi_i + \phi_j - \phi_k - \phi_l),
    \end{aligned}
\end{align}
where $e_{ij} = 1$ if $i = j$ and 2 otherwise.

The system of Zakharov equations written in terms of the amplitudes and phases is obtained from Hamilton's equations
\begin{align}
\begin{aligned}
    &\frac{d|b_i|^2}{dt} = -\frac{\partial H}{\partial \phi_i},\\
    &\frac{d\phi_i}{dt} = \frac{\partial H}{\partial |b_i|^2},
\end{aligned}\quad\quad\text{for $i = 1\ldots,N$.}
\end{align}

Let us tackle the most general case of four waves satisfying the wavenumber resonance condition $\bk_1 + \bk_2 = \bk_3 + \bk_4$. The discrete Hamiltonian \eqref{eq:Hamiltonian discrete} becomes
\begin{align} \label{eq:Hamiltonian-action-angle}
    H = \sum_{i=1}^4 \omega_i|b_i|^2 + \frac{1}{2}\sum_{i,j=1}^4 e_{ij}T_{ij}|b_i|^2|b_j|^2 + 4 T_{1234}  \sqrt{|b_1|^2 |b_2|^2 |b_3|^2 |b_4|^2}  \cos(\phi_1 + \phi_2 - \phi_3 - \phi_4),
\end{align}
and the corresponding equations of motion are 
\begin{align} \label{eq:class Ia db12dt} 
    \frac{d|b_{1,2}|^2}{dt} &= -\frac{\partial H}{\partial \phi_{1,2}} = 4T_{1234} \sqrt{|b_1|^2 |b_2|^2 |b_3|^2 |b_4|^2}  \sin(\phi_1 + \phi_2 - \phi_3 - \phi_4),\\ \label{eq:class Ia db34dt}
    \frac{d|b_{3,4}|^2}{dt} &= -\frac{\partial H}{\partial \phi_{3,4}} = -4T_{1234} \sqrt{|b_1|^2 |b_2|^2 |b_3|^2 |b_4|^2}  \sin(\phi_1 + \phi_2 - \phi_3 - \phi_4),\\
    \frac{d\phi_i}{dt} &= \frac{\partial H}{\partial |b_{i}|^2} = \omega_i + \Gamma_i + 2T_{1234}\frac{\sqrt{|b_1|^2 |b_2|^2 |b_3|^2 |b_4|^2}}{|b_i|^2}\cos(\phi_1 + \phi_2 - \phi_3 - \phi_4),\quad\text{for $i = 1,2,3,4$.}\label{eq:class Ia individual phases}
\end{align}
Note that when treating the degenerate quartet $2\bk_1 = \bk_2 + \bk_3$ which gives rise to the Benjamin-Feir instability (see Section \ref{ssec:BFI}), the coefficient of \eqref{eq:class Ia db34dt} is -2 rather than -4. This is because there are fewer possibilities to assemble the resonant wavenumbers needed. Otherwise the system essentially changes by equating indices 1 and 2, see Section 2.3 of \cite{Andrade2023instability} for details. The frequency correction terms of wave $i$ are collected in the coefficient $\Gamma_i$, which is
\begin{equation} \label{eq:def-Gamma}
\Gamma_i = T_i |b_i|^2 + 2 \sum_{j\neq i}T_{ij}|b_j|^2.
\end{equation}

The equations of motion form a system of eight coupled ODEs for the amplitudes and phases. 
The structure of the Hamiltonian \eqref{eq:Hamiltonian-action-angle} (or a look at \eqref{eq:class Ia db12dt} and \eqref{eq:class Ia db34dt}) makes clear that the so-called Manley-Rowe relations hold: 
\begin{equation}
|b_1|^2 - |b_1(0)|^2 = |b_2|^2 - |b_2(0)|^2 = -|b_3|^2 + |b_3(0)|^2 = -|b_4|^2 + |b_4(0)|^2.
\end{equation}
Therefore, as long as the initial amplitudes $|b_i(0)|$ are given, the evolution of the amplitudes requires only a single equation. Moreover, the fact that the phases appear only as a single \textit{dynamic phase} variable $\theta=\phi_1 + \phi_2 - \phi_3 - \phi_4$ suggests combining \eqref{eq:class Ia individual phases} for $\phi_1, \, \phi_2, \, \phi_3$ and $\phi_4$ into a single equation for $d \theta/dt.$ 

In fact, at this point it is possible to follow the steps of \cite[Section 6]{Bretherton1964a} and integrate the system in terms of Jacobian elliptic functions. This leads (for the generic quartet) to the fourth-order polynomial found in \cite[Eq.\ (3.9)]{Stiassnie2005}. The dependence of roots of such polynomials on their coefficients, particularly when these coefficients in turn depend on the initial data, the interacting wavenumbers, and the interaction kernel $T_{1234}$ is rather complicated. A simpler alternative, which provides immediate insight and exploits the structure of the problem, is to find suitable auxiliary variables.

A first step in this procedure is to recast the conservation laws \eqref{eq:invariants} and the resonance condition in matrix form 
\[
    \begin{bmatrix}
        1    & 1   & 1   & 1     \\
        k_1  & k_2 & k_3 & k_1+k_2-k_3 \\
        l_1  & l_2 & l_3 & l_1+l_2-l_3 
    \end{bmatrix}
    \begin{bmatrix}
        |b_1(t)|^2\\
        |b_2(t)|^2\\
        |b_3(t)|^2\\
        |b_4(t)|^2
    \end{bmatrix}
    =
    \begin{bmatrix}
        A_{}\\
        M^x_{}\\
        M^y_{}
    \end{bmatrix}, \]
where we write $\bk_i = (k_i,l_i)$ and $M=(M^x,M^y)$ for the components of these 2D vectors. Since $A$ and $M$ are conserved quantities we may replace them by their values at $t=0.$ If we are in a nondegenerate situation such that $\bk_i \neq \bk_j$ for $i\neq j$ the (constant) coefficient matrix on the left has rank 3, and the general solution to this system of equations is given by the sum of any particular solution and a (time-dependent) multiple of $(1,1,-1,-1)$ which spans the kernel of the matrix.

The selection of the particular solution is driven by the physical configuration of interest: if we are interested in a configuration of energy transfer starting from mode $\bk_1,$ we would pick $A(0)=|b_1(0)|^2$ and $M(0)=(|b_1(0)|^2k_1,|b_1(0)|^2 l_1),$ with particular solution $(|b_1(0)|^2,0,0,0)$ and general solution $(|b_1(0)|^2+f(t),f(t),-f(t),-f(t)).$ On the other hand, if we are interested in energy transfer from one bichromatic wave train $\bk_1, \, \bk_2$ to another $\bk_3, \, \bk_4$ the corresponding general solution is $(|b_1(0)|^2+f(t),|b_2(0)|^2+f(t),-f(t),-f(t)).$ This provides a natural auxiliary variable for the problem.

Let us assume for simplicity that $|b_1(0)|^2=|b_2(0)|^2=A/2,$ so that 
\begin{align*}
|b_1(t)|^2 &= |b_2(t)|^2 = A \eta(t),\\
|b_3(t)|^2 &= |b_4(t)|^2 = A (1/2 - \eta(t)).
\end{align*}
This is an initial condition wherein the wave action is equipartitioned among the two modes of a bichromatic sea. In terms of the auxiliary variable $\eta$ and the dynamic phase $\theta$ the Hamiltonian can be written 
\begin{align}\label{eq:Htilde:1}
    \Tilde{H}(\eta,\theta) = -2AT_{1234}\eta(1 - 2\eta)\cos(\theta) - (\Delta_{12}^{34} + A\Omega_0)\eta -\frac{A\Omega_1}{2}\eta^2,
\end{align}
and the equations of motion 
\begin{align} \label{eq:Type Ia d eta d t}
    \frac{d\eta}{dt} & = 2AT_{1234}\eta (1 - 2\eta)\sin(\theta),\\ \label{eq:Type Ia d theta d t}
    \frac{d\theta}{dt} & = \Delta_{12}^{34} + A\Omega_0 + A\Omega_1\eta + 2AT_{1234}(1 - 4\eta)\cos(\theta),
\end{align}
with $\Delta_{12}^{34} = \omega_1 + \omega_2 - \omega_3 - \omega_4,$ and 
\begin{align*}
    \Omega_0 &= T_{13} + T_{23} - T_{33}/2 - 2T_{43} + T_{14} + T_{24}  - T_{44}/2,\\
    \Omega_1 &= (T_{11} + 4T_{12} + T_{22}) + (T_{33} + 4T_{34} + T_{44}) - 4(T_{13} + T_{14} + T_{23} + T_{24}).
\end{align*}

\begin{figure}
\centering
\includegraphics[width=0.35\linewidth]{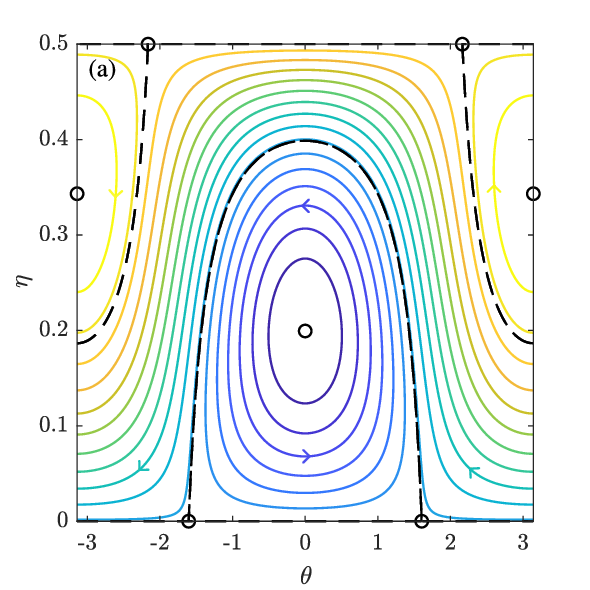}
\includegraphics[width=0.35\linewidth]{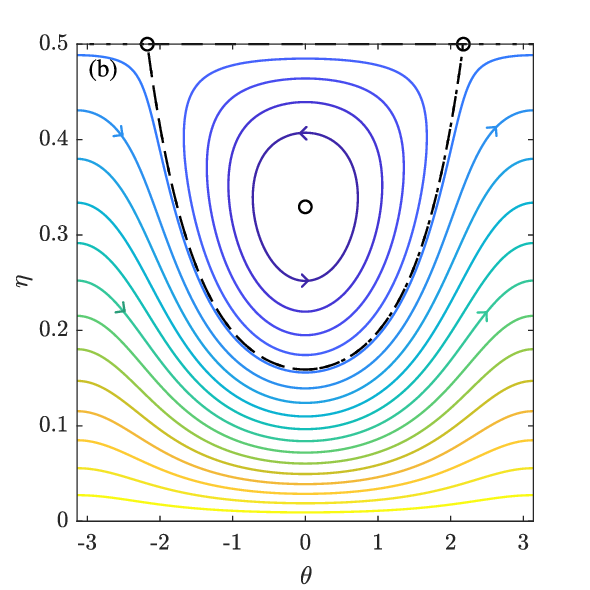}\\
\includegraphics[width=0.35\linewidth]{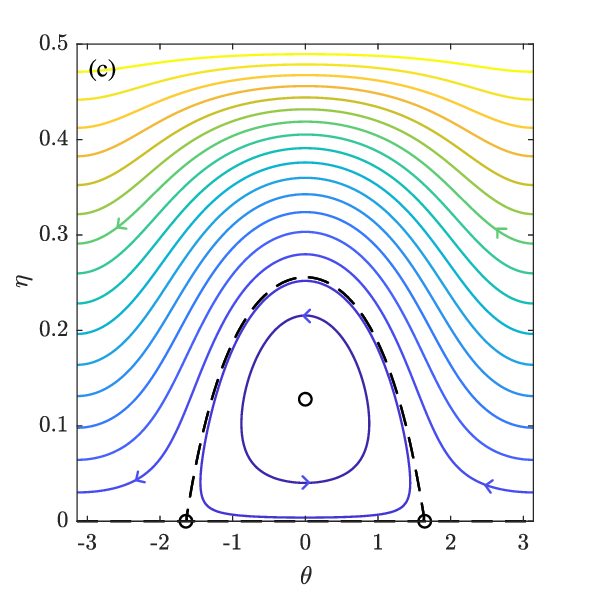}
\includegraphics[width=0.35\linewidth]{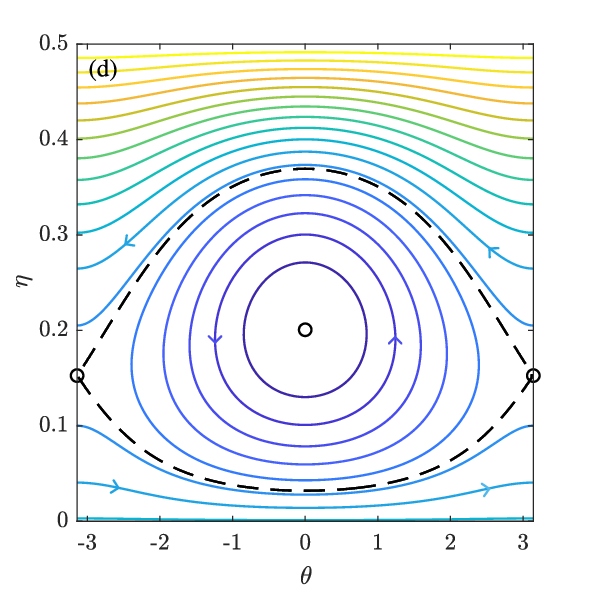}
\caption{Configurations of saddle points and centres (black circles) and separatrices (dashed curves) in the interaction of two bichromatic wave trains with initially equal wave action, described by the dynamical system \eqref{eq:Type Ia d eta d t}--\eqref{eq:Type Ia d theta d t}.}
\label{fig:Phase portraits quartet}
\end{figure}

The advantages of a planar, Hamiltonian description of this four-wave interaction are manifold. The Hamiltonian $\tilde{H}$ has Jacobi matrix 
\begin{equation} J = \begin{pmatrix} \tilde{H}_{\eta \theta} & \tilde{H}_{\theta\theta} \\
-\tilde{H}_{\eta \eta} & -\tilde{H}_{\eta \theta} \end{pmatrix}, \end{equation}
whose trace vanishes and whose determinant is $\det(J) = \tilde{H}_{\theta \theta} \tilde{H}_{\eta \eta} - \tilde{H}^2_{\eta \theta}.$ Thus the eigenvalues are $\lambda_{1,2} = \pm \sqrt{-\det(J)}$ and either purely real or purely imaginary, so that fixed points are saddles and centres only. The phase space is $\{(\eta,\theta)\in [0,1/2]\times[-\pi,\pi]\},$ with orbits coinciding with the level lines of the Hamiltonian. A few such orbits are shown in Figure \ref{fig:Phase portraits quartet}, which depicts four generic configurations of fixed points and separatrices.

\begin{figure}
\centering
\includegraphics[width=0.9\linewidth]{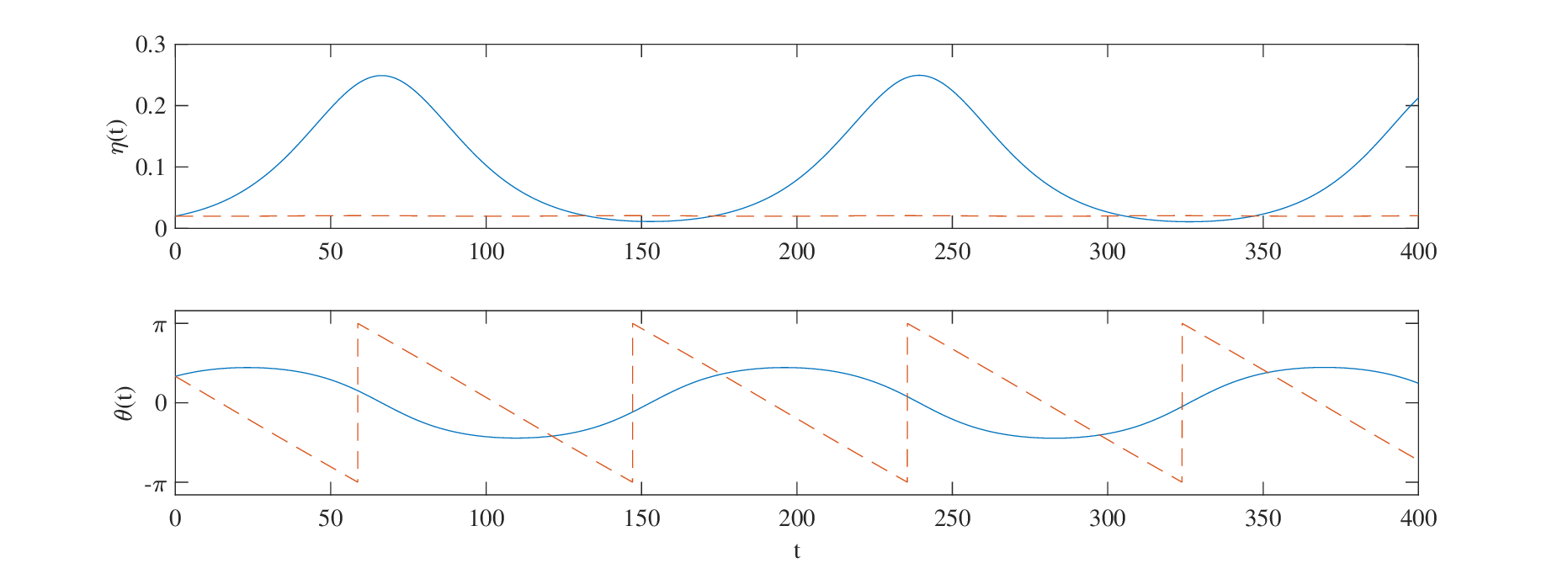}
\caption{Plot of $\eta$ and $\theta$ with time for two distinct cases. Blue curves: a quartet with initial conditions surrounding the centre at $\theta=0$ in panel (a) Figure \ref{fig:Phase portraits quartet}. Red, dashed curves: a quartet with initial conditions towards the bottom ($\eta=0.02$) of the phase space shown in panel (b) of Figure \ref{fig:Phase portraits quartet}. The former shows energy exchange and the attendant effect on the dynamic phase. The latter shown no energy exchange, with dynamic phase essentially a linear function of time.}
\label{fig:Quartet eta-theta with t}
\end{figure}

It remains to interpret the dynamics of this system, which are rich and revealing. Shown in Figure \ref{fig:Quartet eta-theta with t} are plots of both $\eta$ -- which represents energy exchange among modes -- and $\theta$ with time. Here only two typical trajectories in the phase plane have been plotted for simplicity: a trajectory surrounding a centre, which is shown in blue, and which exhibits energy exchange along with concomitant change in the dynamic phase, which remains confined to a narrow range $\theta \in (-1.4,1.4)$. On the other hand, the red (dashed) curve shows a case where almost no energy exchange takes place. This can be engendered by the separation of the Fourier modes, by the relative inclinations (angles) of the waves, or by the slopes of the waves involved. The consequence is that the waves behave almost like linear waves, despite the fact that we are considering a solution to the third-order nonlinear problem.

Of course there are other cases which we can find via phase-plane analysis: the fixed points, shown as black circles in Figure \ref{fig:Phase portraits quartet}, are quartets of waves whose dynamic phase as well as amplitudes are all constant. In such a case the wave field is in a steady-state, and effectively indistinguishable from a linear wave field, except for the different phase velocities. The heteroclinic orbits form breather solutions with four modes, which tend asymptotically (as $t\rightarrow \pm \infty$) to a bichromatic background. Such cases have been recently considered \cite{Andrade2023,Andrade2023instability} and provide a new perspective on an old problem. Indeed, we can use this to appreciate that, while the ``generic" long-time evolution of Benjamin-Feir instability is indeed a type of Fermi-Pasta-Ulam recurrence (as pointed out by \cite[Sec.\ VII.C]{Yuen1982}),  particular initial conditions may give rise to a single modulation or even to steady states.

\section{Dispersion corrections}
\label{sec:Dispersion corrections}

One particularly useful property of the Zakharov reformulation of the water wave problem is that it lays bare the various contributions to the behaviour of waves up to a specified order of nonlinearity. We have chosen from the outset to restrict ourselves to the case of cubic nonlinearity, which captures much of the interesting behaviour found in deep water waves. Having discussed, in Section \ref{sec:Discrete interactions}, the importance of energy exchange, here we will take the opposite tack and assume it is absent entirely.

What is meant by this is best seen by looking at the system in terms of amplitude and phase separately, i.e.\ the action angle variables shown in \eqref{eq:class Ia db12dt}--\eqref{eq:class Ia individual phases}. More generally, for an arbitrary number of discrete modes we can write the equation for the phase 
\begin{equation}
\frac{d\phi_n}{dt} = \omega_n + \Gamma_n + |b_n|^{-1} \sum_{p} \sum_{q \neq n} \sum_{r \neq n} T_{npqr} \delta_{np}^{qr} |b_p||b_q||b_r| \cos(\theta_{npqr})
\end{equation}
where $\Gamma_i$ is defined in \eqref{eq:def-Gamma}. In this formulation we see that if there are no nontrivial resonances which satisfy the Kronecker delta $\delta_{np}^{qr}$, the evolution of the phase is simply 
\[ \phi_n = \Omega_n t = \left( \omega_n + \Gamma_n \right)t = \left( \omega_n + T_n |b_n(0)|^2 + 2 \sum_{j\neq n}T_{nj}|b_j(0)|^2\right) t. \]
Here we recognise the nonlinear dispersion corrections: the linear frequency $\omega_n$ is shifted due to the self-interaction found in the Stokes wave (Section \ref{ssec:Stokes waves}) and the mutual interaction found in the bichromatic wave (Section \ref{ssec:Bichromatic wave trains}), in terms of the initial amplitudes $|b_i(0)|,$ which are related to the measured Fourier amplitudes by \eqref{eq:B-to-a}. The kernels appear initially to be intimidating, but only symmetric kernel expression appear. If the waves are unidirectional these are extremely simple, and for multidirectional waves the formulas given in Appendix \ref{sec:Compact kernels} nevertheless allow for straightforward computational implementation.

Even if resonances which satisfy $\delta_{np}^{qr}$ cannot be ruled out, the dispersion correction $\Omega_n = \omega_n + \Gamma_n$ is still useful. This is because, in realistic sea-states -- simulated with many modes-- the individual Fourier amplitudes oscillate in a non-recurrent fashion. The amplitudes can thus be treated as constant to a first approximation, and the integration of $d\phi_n/dt$ interpreted in a suitable averaged sense. This can be checked by Monte-Carlo simulations of the Zakharov equation, see \cite[Appendix A]{Stuhlmeier2019}. It is also borne out by the utility of this very dispersion correction in producing accurate, deterministic forecasts of waves, both from laboratory measurements \cite{Galvagno2021} and from synthetic data generated from realistic simulations of sea states \cite{Stuhlmeier2021,meisner2023}.

For a continuous spectrum $\Psi(\bk)$ it can be assumed that the number of modes $N$ tends to infinity.
The limit of a continuous wave-number energy spectrum $\Psi(\bk)$ is approached by considering a square grid of wave numbers with spacing $d \bk,$ so that 
\begin{equation} \label{eq: amplitude wavenumber spectr} {a_n^2}/{2} = \Psi_n = \Psi(\bk_n) d \bk. \end{equation}
Taking the limit $d\bk \rightarrow 0,$ the nonlinear corrected frequency can then be rewritten as
\begin{align} \label{eq: Nonlinear Frequency}
{ \Omega(\bk) = \omega(\bk) + 8 \pi^2 \int T(\bk,\bk_1,\bk,\bk_1) \frac{\omega(\bk_1)}{|\bk_1|} \Psi(\bk_1) d\bk_1.} 
\end{align}

\begin{figure}[ht!]
\centering
\includegraphics[width=0.9\linewidth]{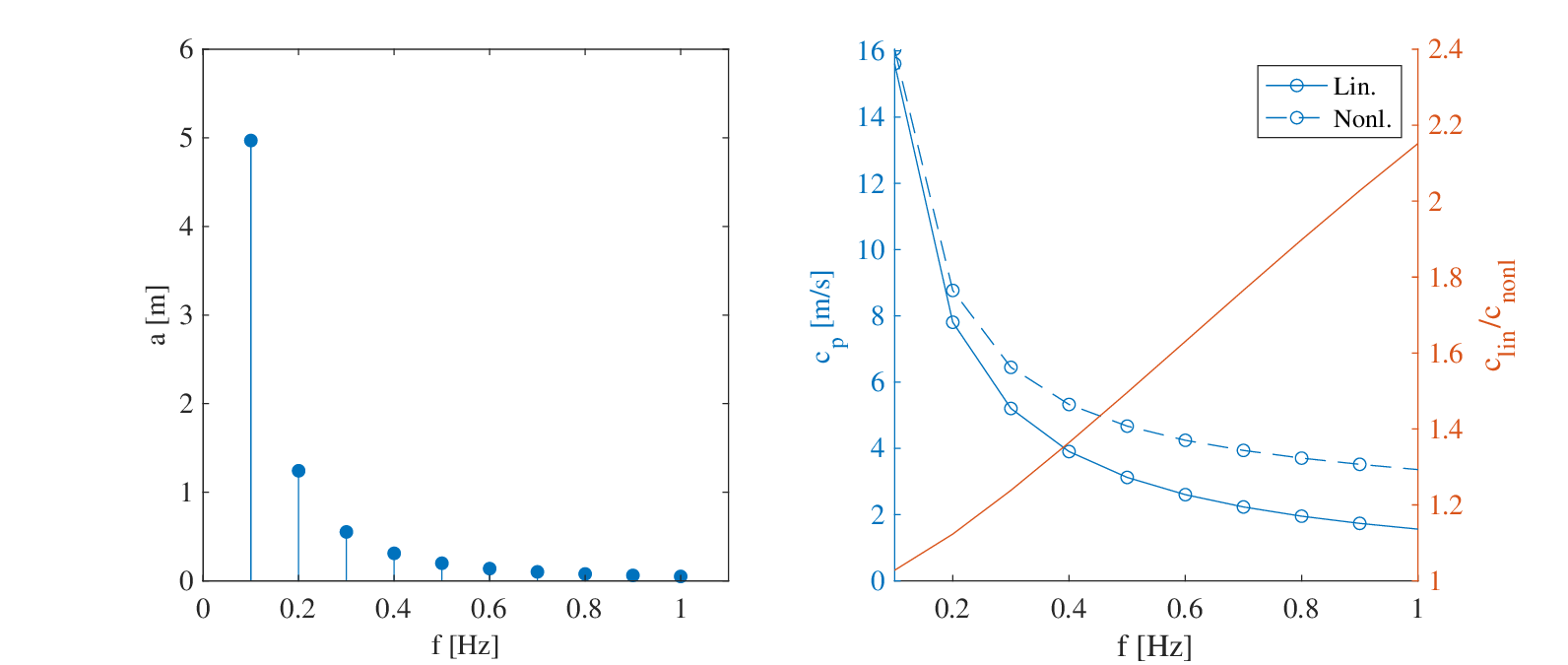}
\caption{(Left panel) Representation of a toy spectrum of ten equally spaced modes from $f=0.1$ to $f=1$ Hz, each mode with steepness $\epsilon = ak=0.2.$ (Right panel) Linear deep-water dispersion relation for each mode (solid curve) and nonlinear corrected dispersion relation \eqref{eq: Nonlinear Frequency} (dashed curve), together with the relative correction (red curve).}
\label{fig:Nonlinear dispersion corrections}
\end{figure}

The effect of dispersion corrections is notably asymmetric, as already evident in the exact, bichromatic solution to the Zakharov equation found by Longuet-Higgins and Phillips (see Section \ref{ssec:Bichromatic wave trains}). There we find that the long waves have a significant effect on the frequency of the short waves, but not vice versa. The dispersion relation is affected by the squared amplitudes of all waves, mediated by the symmetric interaction kernel. This is shown in Figure \ref{fig:Nonlinear dispersion corrections}, where a simple spectrum of ten modes, each with steepness $\epsilon = 0.2$ is used to calculate the linear and nonlinear phase velocities. For the shortest waves, the difference between the two phase velocities is more than a factor of two, while for the longest waves the difference is nearly negligible (they fail to ``feel" the short waves at all).

\begin{figure}[h!]
\centering
\includegraphics[width=0.9\linewidth]{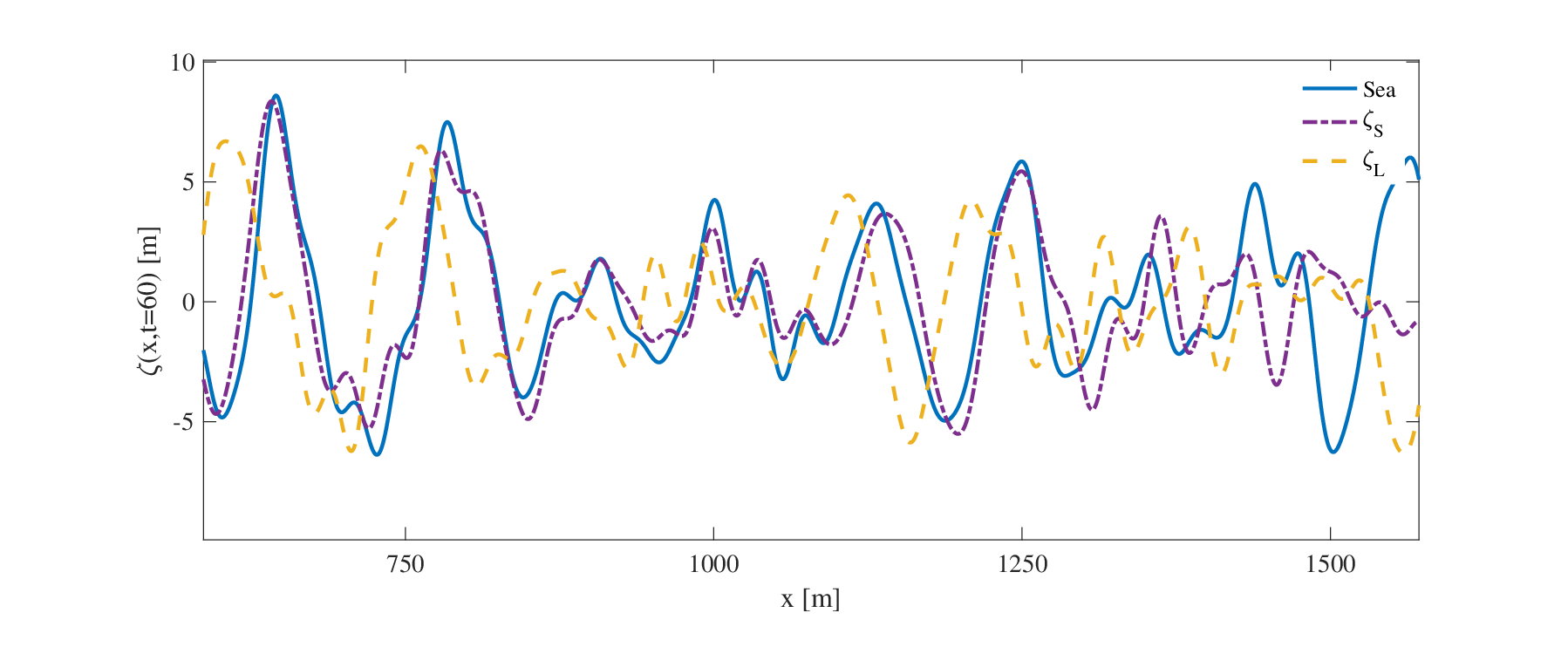}
\caption{An example of the effect of nonlinear frequency correction on simulated waves. A realisation of a (JONSWAP) spectrum (spectral peak $k_p = 0.05$ 1/m, $\gamma=5$) is generated at $t=0$ s, and modes are propagated forward to $t=60$ s using either linear (yellow, dashed curve) or nonlinear dispersion (purple, dash-dotted curve). The actual simulated sea is shown in blue. The nonlinear dispersion correction is enhanced by the (artificially) high steepness of $\epsilon=0.2$}
\label{fig:Nonlinear forecast}
\end{figure}

The effects of such dispersion corrections are important when simulating or forecasting waves. Indeed, while each Fourier mode simply undergoes a phase correction, the sum of all those phase corrections has a dramatic effect on the overall water surface. Such a case is shown in Figure \ref{fig:Nonlinear forecast}. Thus, while nonlinear interactions (and the eventual chaotisation which can be expected from interactions with many modes, see \cite{Annenkov2001}) occur on a slow time scale $O(\epsilon^2)$, the changes in phase velocity due to nonlinear corrections lead to significant differences already on a timescale of $O(\epsilon),$ see the discussion in \cite{Stuhlmeier2021}.

These formulations can be used for deterministic forecasting of waves, on the basis of initially measured Fourier amplitudes (which enter into the $|B_i(0)|$), and comparisons with HOS and wave flume experiments have shown that such forecasts are a significant improvement over linear theory at no additional computational cost \cite{Stuhlmeier2019, Stuhlmeier2021, Galvagno2021, meisner2023}. Such studies have been undertaken in infinite water depth only.
The symmetric Zakharov kernels in finite depth exhibit singularities, and therefore require careful treatment.

\section{Conclusions \& Discussion}

In the preceding sections we have introduced the Zakharov equation and demonstrated its utility in a variety of contexts. It springs forth from the Hamiltonian for the water wave problem, and the idea of reducing the Hamiltonian via a series of transformations that eliminate non-resonant contributions. In principle the equation can also be derived from the governing equations directly via a perturbative approach, but the lack of a Hamiltonian structure makes it difficult to correctly account for all the symmetries of the problem. This issue was resolved definitively in the seminal contribution of Krasitskii \cite{Krasitskii1994}. The Zakharov equation is a mother equation to the NLS \cite{Zakharov1968} and higher order NLS equations \cite{Stiassnie1984c}, but is much less known and considerably less popular, despite the inherent advantage of possessing no additional restriction on bandwidth. In contrast, the NLS is derived from the mother equation by assuming all wavenumbers are close to a central wavenumber, and inserting a Taylor expansion of the frequency, as well as approximating the interaction kernels.

We have focused exclusively on the infinite-depth Zakharov equation. This is despite the equation being derived for a flat bed at arbitrary depth by Krasitskii \cite{Krasitskii1994}. Nevertheless, for several years workers in the field have been aware of difficulties in the finite-depth kernels for the Zakharov formulation. These were pointed out by Janssen and Onorato \cite{Janssen2007}, who treated the case of apparent singularities in the self-interaction kernels $T_{aaaa}$. Stiassnie \& Gramstad  \cite{Stiassnie2009b} later investigated the bichromatic kernels $T_{abab}$ as well, splitting the expressions into regular and singular parts. Several years later,  Gramstad \cite{Gramstad2014} rederived the Zakharov equation in finite depth along the lines of Krasitskii \cite{Krasitskii1994}, albeit with explicit terms for the mean flow and mean surface level.
In a recent incisive analysis, Pezzutto and Shrira \cite{Pezzutto2023} have shown that these singularities can be eliminated by correct treatment of Dirac delta functions appearing in the formulation. They also highlight some of the differences between the strictly 1D and 2D formulations of the problem, and suggest that the former may require further attention.

Our focus has been on uses of the Zakharov equation for the description of deterministic waves. This fails to mention the very important role that this equation plays in the derivation of the kinetic equation which is used for stochastic wave forecasts. The reader interested in this aspect of the problem is directed to the excellent book by Janssen \cite{Janssen2004}, or for a very brief account, to Chapter 14.10 of Mei \textit{et al} \cite{Mei2005}. The Zakharov equation has been used to compare with the kinetic equation via direct numerical simulation (essentially Monte-Carlo simulation) \cite{Annenkov2001a,Annenkov2006}, and novel wave-kinetic equations have been derived from this starting point in recent years from the generalised kinetic equation of Annenkov \& Shrira \cite{Annenkov2016}, the modified kinetic equation of Gramstad \& Stiassnie \cite{Gramstad2013, Gramstad2016}, to the inhomogeneous equations considered by Crawford et al \cite{Crawford1980} and recently investigated by Stiassnie and others \cite{Stuhlmeier2017,Stuhlmeier2019a,Andrade2020}.

We have mentioned the link between the Zakharov formulation and the very successful higher order spectral method (HOS) (see Chapter 15 of Mei \textit{et al} \cite{Mei2005}). Indeed, Tanaka \cite{Tanaka2001} has shown that the Zakharov equation is equivalent to the HOS for the same order of nonlinearity, while being computationally considerably more efficient. Thus while the Zakharov equation has an analytically useful structure, and its discretised form boils down simply to a system of coupled, nonlinear ODEs, it is not the most efficient tool for numerical modelling of water waves. An alternative formulation using the free surface envelope (instead of the free surface itself) and the velocity potential at the free surface as canonical variables has been recently developed by \cite{Li2023}. This effectively develops an NLS-like Hamiltonian formulation, which gives direct access to the envelope evolution, and which may prove a useful tool for further computational implementation.

In flume experiments, as mentioned previously, the evolution of the waves is typically measured with gauges which are positioned at different locations along the length of the flume. This means that each gauge is recording a time series, and what can be measured is the evolution in space from one gauge to the next. This contrasts with the description of the Zakharov equation presented here, where the assumption is that the domain is homogeneous in space (i.e.\ described by a wavenumber spectrum) and evolving in time. This requires a reformulation of the equation into the so-called spatial Zakharov equation, developed by Shemer and co-workers in the early 2000s \cite{Shemer2001,Shemer2002} and subsequently used to interpret flume experiments \cite{Shemer2017}.
Many of the developments we have presented can be rederived for the spatial equation, for example the idea of a nonlinear correction to the phase velocity. In the spatial equation the frequency is fundamental, so the wavenumber changes accordingly as
\begin{equation} \label{eq: wavenumber correction}
K_n = k_n - \frac{1}{c_{g,n}} \Gamma_n. 
\end{equation}
In both the spatial and temporal case, the symmetric Zakharov kernels $T_{npnp}$ as functions of wavenumber are used, and $K_n$ is the corrected wavenumber, with $\Gamma$ defined in \eqref{eq:def-Gamma}. Such a nonlinear wavenumber correction has been used successfully in testing simple deterministic wave forecasting methods by Galvagno et al \cite{Galvagno2021}.

Hamiltonian formulations have in recent years been extended to a diverse variety of contexts in water waves, including waves with vorticity \cite{Constantin2005}, stratified rotational flows \cite{Constantin2016}, rotational capillary waves \cite{Martin2016}, and even internal waves with variable topography \cite{Ivanov2022}. All of these are scenarios of great practical interest, and this raises the question whether transformations to a Zakharov-type equation, which adeptly eliminates the non-resonant contributions and lays bare the essential energy transfer mechanisms, can be found when starting from such cases. 

\section*{Acknowledgements}
The author is grateful for the hospitality of University College Cork during the workshop \textit{Nonlinear Dispersive Waves} held in April 2023. Debbie Eeltink kindly shared the experimental data shown in Figure \ref{fig:BFI}, and Christopher Luneau and Hubert Branger from the IRPHE/Pytheas Aix Marseille University are thanked for their help with the underlying experiments. Particular gratitude is owed to Michael Stiassnie, who introduced the author to the Zakharov formulation and has been a valued collaborator and mentor for the past decade. 

\appendix

\section{Compact formulations of the kernels}
\label{sec:Compact kernels}

The kernels of the Zakharov formulation encode essentially all the information about the water wave problem. For the free wave part only a single kernel, which we denote by $T(k_1,k_2,k_3,k_4)$ is needed. Expressions for this can be found in Krasitskii \cite{Krasitskii1994}, Chapter 14 of Mei et al \cite{Mei2005}, or Janssen \& Onorato \cite{Janssen2007}. This kernel moreover has several important symmetries which follow from the Hamiltonian structure:
\[ T(k_1,k_2,k_3,k_4) = T(k_2,k_1,k_3,k_4) = T(k_1,k_2,k_4,k_3) = T(k_3,k_4,k_1,k_2). \]
The easy way to remember these symmetries is to think about the permutations that leave the resonance condition $k_1 + k_2 = k_3 + k_4$ invariant. For the bound wave components further kernels are necessary -- these can also be found in the references quoted above.

For unidirectional waves in deep water, we can find a convenient condensed expression for $T_{abcd}$ in Kachulin et al \cite{Kachulin2019} (adjusted by a factor of $2 \pi$ due a different convention for the Fourier transform). The kernels can be further simplified by making use of the homogeneity property $T(\alpha k_a, \alpha k_b, \alpha k_c, \alpha k_d) = \alpha^{3} T(k_a,k_b,k_c,k_d).$ 

For the case of unidirectional waves in deep water ($k_i>0$ for all $i$), considerable simplifications to the kernels are thus possible.
\begin{align} \label{eq:Kernel-unidir-one-wave}
    T(k,k,k,k)&=\frac{k^3}{4 \pi^2}\\ \label{eq:Kernel-unidir-two-wave}
    T(k_a,k_b,k_a,k_b)&=\frac{1}{4 \pi^2} k_a k_b \min(k_a,k_b)\\ \nonumber
    T(k_a,k_b,k_c,k_d)&=\frac{(k_a k_b k_c k_d)^{1/4}}{32 \pi^2} \left[ \left(k_a k_b\right)^{1/2} + \left( k_c k_d \right)^{1/2} \right]\\
    &\cdot \left( k_a + k_b + k_c + k_d - \left[ |k_a - k_c| + |k_a-k_d| + |k_b-k_c| + |k_b-k_d| \right] \right)
\end{align}
For a degenerate quartet of waves $2k_a=k_b + k_c,$ which appears in the 1D Benjamin-Feir instability (Section \ref{ssec:BFI}) this enables us to write the kernel
\begin{align}
    T(k_a,k_a,k_b,k_c) = \frac{1}{8\pi^2}\left[\min(k_b,k_c)(k_a^2k_bk_c)^{1/4}(k_a + \sqrt{k_bk_c})\right].
\end{align}
For non-unidirectional cases, the kernels are more involved. There is a relatively compact expression 
for the symmetric interaction kernel $T(\bk_i,\bk_j,\bk_i,\bk_j)$  given by 
\begin{align} \nonumber
& T(\bk_i, \bk_j, \bk_i, \bk_j)  = - \frac{1}{16 \pi^2 (|\bk_i| |\bk_j|)^{1/2}} \left[ 3(|\bk_i| |\bk_j|)^2 \right. \\ \nonumber
& \left. + (\bk_i \cdot \bk_j)(\bk_i \cdot \bk_j - 4 (|\bk_i| + |\bk_j|)(|\bk_i| |\bk_j|)^{1/2}) \right. \\
& + \frac{2(\omega_i - \omega_j)^2 (\bk_i \cdot \bk_j + |\bk_i| |\bk_j|)^2}{g |\bk_i - \bk_j| - (\omega_i - \omega_j)^2} 
+ \left. \frac{2(\omega_i + \omega_j)^2 (\bk_i \cdot \bk_j - |\bk_i| |\bk_j|)^2}{g |\bk_i + \bk_j| - (\omega_i + \omega_j)^2} \right],
\end{align}
see \cite{Leblanc2009} or substitute \eqref{eq: Linear Dispersion Relation} into (3.9b) of \cite{Stiassnie2009b}.

\newcommand{\etalchar}[1]{$^{#1}$}

\end{document}